\documentclass[twocolumn, twocolappendix]{aastex631}
\usepackage[utf8]{inputenc}
\usepackage{graphicx}
\usepackage{color}
\usepackage{here}
\usepackage{ascmac}
\usepackage{fancybox}
\usepackage{bm}
\usepackage{amsmath}
\usepackage{amssymb}
\usepackage{mathrsfs, mhchem}
\usepackage[T1]{fontenc}

\newcommand{\Myr}{\mathrm{\,Myr}}
\newcommand{\Msun}{M_\odot}
\newcommand{\yr}{\mathrm{\,yr}}
\newcommand{\unit}[1]{\mathrm{\,#1}}

\newcommand{\Mdisk}{M_\mathrm{disk}}
\newcommand{\Macc}{\dot{M}_\mathrm{acc}}

\begin{document}

\title{Secret of Longevity: Protoplanetary Disks as a Source of Gas in Debris Disks}

\author[0009-0000-7797-8347]{Wataru Ooyama}
\affiliation{Department of Physics, Graduate School of Science, Kyoto University, Sakyo, Kyoto 606-8502, Japan}
\email{ooyama.wataru.64z@st.kyoto-u.ac.jp}

\author[0000-0002-1803-0203]{Riouhei Nakatani}
\affiliation{Dipartimento di Fisica, Universit\`a degli Studi di Milano, Via Celoria, 16, I-20133 Milano, Italy}

\author[0000-0003-3127-5982]{Takashi Hosokawa}
\affiliation{Department of Physics, Graduate School of Science, Kyoto University, Sakyo, Kyoto 606-8502, Japan}

\author[0000-0003-1050-3413]{Hiroto Mitani}
\affiliation{Faculty of Physics, University of Duisburg-Essen, Lotharstra{\ss}e 1, D-47057 Duisburg, Germany}
\affiliation{Department of Physics, School of Science, The University of Tokyo, 7-3-1 Hongo, Bunkyo, Tokyo 113-0033, Japan}

\author[0000-0001-8292-1943]{Neal J. Turner}
\affiliation{NASA Jet Propulsion Laboratory, California Institute of Technology, 4800 Oak Grove Dr, Pasadena, CA 91109, USA}

\begin{abstract}
While protoplanetary disks (PPDs) are generally thought to 
disperse within several million years, recent observations have revealed gas in their older counterparts, debris disks. The origin of this gas remains uncertain, with one possibility being the unexpectedly long survival of PPDs (the primordial-origin scenario). 
To explore the plausibility of this scenario, we conduct 1D disk evolution simulations, varying parameters like stellar mass, disk mass, turbulent stress, and the model of magnetohydrodynamic winds, while incorporating stellar evolution to account for time-varying photoevaporation rates. Our focus is on disks where small grains are depleted, as these are potentially long-lived due to reduced far-ultraviolet photoevaporation.
Our results show that gas in these disks can survive beyond $10\Myr$ regardless of the stellar mass, provided they are initially massive ($\Mdisk\approx 0.1M_*$) with relatively weak turbulent stress ($\alpha \ll 10^{-2}$). The longest lifetimes are consistently found for $M_* = 2\Msun$ across a wide parameter space, with gas typically persisting at $\sim 10$--$10^3\unit{au}$. Roughly estimated CO masses for these disks fall within the observed range for the most massive gas-rich debris disks around early A~stars. These alignments support the plausibility of the primordial-origin scenario. Additionally, our model predicts that accretion persists for as long as the disk survives, which could explain the accretion signatures detected in old disks hosted by low-mass stars, including Peter Pan disks. Our finding also suggests that ongoing accretion may exist in gas-rich debris disks. Thus, 
searching for accretion signatures could be a key to determining the origins of gas in debris disks.
\end{abstract}

\section{Introduction}
\label{Intro}
Since protoplanetary disks are the birthplace of planets, understanding their evolution is crucial for constructing a theory to explain the diversity of the more than 5,000 exoplanets \citep{exoplanet_archive} discovered to date \citep{Venturini_2020,Drazkowska_2023,Burn_2024} .
One of the important parameters in planet formation theory is gas disk lifetime, as it sets the time limit of gas giant formation and disk-planet interactions. Disk lifetimes are typically estimated through near- or mid-infrared emissions from inner hot dust \citep[e.g.,][]{Haisch2001, Meyer2007, Hernandez2007, Mamajek2009, Ribas2014, Ribas2015} and UV or H$\alpha$ emissions from accreting gas \citep[e.g.,][]{Kennedy2009, Fedele2010, Sicilia-Aguilar2010}. Overall, these studies have classically suggested that protoplanetary disks 
disperse within about $10\Myr$.
We note that these observations probe the existence of inner disks ($\lesssim 1$--$10\unit{au}$). 

Disk dispersal primarily occurs in two ways: accretion and winds.  Accretion onto the star results from both angular momentum transfer within the disk by turbulent stresses and angular momentum extraction from the disk by magnetohydrodynamical (MHD) disk winds.  The MHD winds and also the winds launched through photoevaporation of the disk surface layers can remove gas to interstellar space.
Several studies have explored long-term disk evolution using 1D secular evolution simulations, including various subsets of these disk-dispersal processes \citep[e.g.,][]{Linden-Bell+1974, 2001_Clarke, Alexander06, Suzuki+10, Morishima12, Suzuki+16, Kunitomo2020, Komaki+2023, Weder+2023, Tong+24,Kunitomo+21,2024_Ronco} and consistently suggest that gas disks essentially  
disperse within about $10\Myr$.

After the dispersal of protoplanetary disks, debris disks are left over as remnants\citep{Ercolano+2017,Elcolano_Pascucci_2017,Michel_2021}. These disks consist of optically thin dust produced by collisions between planetesimals, making them secondary in nature. 
Traditionally, debris disks were thought to be devoid of gas, but recently, approximately 20 debris disks with gas components have been discovered \citep[gaseous debris disks; e.g.,][]{Kospal2013, Marino2016, Lieman-Sifry2016, Moor2017, Hughes2017, Higuchi2017, Higuchi2019a, Higuchi2019b}.

There are two scenarios proposed to explain the unknown origin of the gas. The first is the primordial-origin scenario, where the gas observed in these disks is a remnant of the original protoplanetary disk \cite{2023_Iwasaki,2022_Smirnov-Pinchukov,Nakatani2021,Nakatani+23}. This implies that some protoplanetary disks must have survived longer than the classical estimates ($\leq 10\Myr$). The second is the secondary-origin scenario, where the gas is released in a similar manner to the dust components after the parental protoplanetary disk has gone.

Historically, constraints from both theoretical models and observations of protoplanetary disk lifetimes have largely ruled out the primordial-origin scenarios, leading to a focus on the plausibility of the secondary-origin scenarios 
\citep{Kral2016, Kral2019, Moor2019, Matra2019, Marino2020, Cataldi2020, Marino2022}. The basic premise of the secondary-origin models is that CO and \ion{C}{1} released from planetesimal collisions accumulate to self-shield against the interstellar and host star's radiation while spreading viscously. The strength of viscosity is parameterized using \citep{Shakura_Sunyaev1973}. \citet{Marino2020} found that high viscosities of $\alpha$$\approx 0.1$ are preferred to match observations through population synthesis. However, \citet{2023_Cataldi} recently demonstrated that these models often overpredict \ion{C}{1} masses. Additionally, \citet{Marino2022} showed that vertical mixing of CO and \ion{C}{1} can shorten CO lifetimes, and if the mixing is strong, the CO gas does not have enough time to spread inside the planetesimal belt. While secondary-origin models are still under development, they have been effective in explaining certain observed characteristics of gaseous debris disks, particularly those with low CO masses \citep[e.g., $\beta$~Pic, HD~181327 and Fomalhaut;][]{Dent+14, Marino2016, Matra+17}, where the secondary-origin scenario is deemed more plausible.

On the other hand, \cite{Nakatani2021} recently revisited the plausibility of the primordial-origin scenarios and proposed a possible pathway for protoplanetary disks to survive beyond $10\Myr$, especially around A~stars. 
A follow-up study by \citet{Nakatani+23}, using a zero-dimensional (0D; one-zone) model, demonstrated that protoplanetary disk lifetimes could extend beyond $10\Myr$ if small grains and polycyclic aromatic hydrocarbons (PAHs) are sufficiently depleted to suppress far-ultraviolet (FUV; $6 \unit{eV} \lesssim h \nu < 13.6 \unit{eV}$) photoevaporation driven by grain photoelectric heating. 
Additionally,  
the low extreme-ultraviolet (EUV; $13.6\unit{eV}< h \nu \lesssim 100\unit{eV}$) and X-ray ($100 \unit{eV} \lesssim h \nu \lesssim 10 \unit{keV}$) luminosities of A~stars further contribute to reducing disk dispersal.  
The weaker EUV and X-ray radiation in A~stars result in systematically longer disk lifetimes compared to low-mass stars, providing a natural explanation for the higher incidence of gaseous debris disks around A~stars. Moreover, it has been shown that the primordial-origin models can reproduce the observed CO and \ion{C}{1} masses within the standard thermochemistry framework \citep{2023_Cataldi, 2023_Iwasaki}. 

In the present study, we extend the one-zone model of \citet{Nakatani+23} to a one-dimensional (1D) secular evolution model. This 1D approach allows us to directly compare the spatial distribution of disk gas and the time evolution of the accretion rate with observations, enabling a more detailed assessment of the plausibility of the primordial-origin scenario.  

The remainder of this paper is structured as follows. In Section~\ref{method}, we describe the numerical methods and the models considered. Our results are presented in Section~\ref{result}. 
Finally, in Sections~\ref{Discussion} and \ref{conclusion}, we provide discussions and concluding remarks.
Throughout the paper, we use the term ``gaseous debris disks'' for debris disks containing gas and ``gas-rich debris disks'' to specifically refer to those with relatively large CO masses ($M_\mathrm{CO} > 10^{-4}\,M_\oplus$). Our main focus is the origin of gas-rich debris disks. However, this distinction does not influence the overall results and conclusions. 

\section{Method}
\label{method}

In this section, we outline our calculation method. In Section~\ref{basic_wq}, we show the basic equations we solve to study the secular evolution of the surface density of the disk, $\Sigma(t,r)$. In Sections~\ref{DW} and \ref{photoeva}, we describe the MHD disk wind 
and photoevaporation models, which provide essential terms driving the disk dispersal. Section~\ref{Temp} summarizes the method used to solve the disk temperature. 
In Sections~\ref{initial_condi} and \ref{disk_model}, we describe the initial and boundary conditions for the models and cases examined.

\subsection{Basic Equations}
\label{basic_wq}

Our basic equation is similar to that considered in recent studies \cite{Kunitomo2020}, \cite{Komaki+2023}, and \cite{Weder+2023}, and it is based on the formula given by \cite{Suzuki+16}:
\begin{multline}
    \frac{\partial \Sigma}{\partial t}=\frac{1}{r} \frac{\partial}{\partial r} \bigg[ \frac{2}{r\Omega} \bigg\{ \frac{\partial}{\partial r} \left( r^{2} \int \mathrm{d}z \left( \rho v_{z}\delta v_{\phi} -\frac{B_{r} B_{\phi}}{4\pi} \right)  \right) \\ +r^{2}\left( \rho \delta v_{\phi} v_{z} -\frac{B_{\phi}B_{z}}{4\pi} \right)_\mathrm{w} \bigg\} \bigg] -\left( \rho v_{z} \right)_\mathrm{w} -\dot{\Sigma}_\mathrm{pw} ,
    \label{eq:basic_eq1}
\end{multline}
where $\Sigma(t,r)$ is the gas surface density, $r$ the distance from the central star, $\Omega$ the Keplerian angular velocity, $v$ the gas velocity, $B$ the magnetic flux density, and $\dot{\Sigma}_\mathrm{pw}(t,r)$ is the local mass loss rate by photoevaporation wind. 
The indices associated with $v$ and $B$ represent the components of the cylindrical coordinates $(r, \phi, z)$.
The velocity difference $\delta v_{\phi}\equiv v_{\phi}-r\Omega$ represents the deviation of the rotational velocity from the background Keplerian value. 
The index "w" associated with the second and third terms on the right-hand side indicates the MHD disk wind. 
The first and second terms on the right-hand side account for mass transport due to the turbulent viscosity and disk wind, respectively. The third and fourth terms represent the mass loss driven by the disk wind and photoevaporation.

We employ the so-called $\alpha$-parameter prescription for modeling the mass transfer through the disk caused by the turbulence \citep{Shakura_Sunyaev1973},  
\begin{equation}
    \int \mathrm{d}z \left( \rho v_{z}\delta v_{\phi} -\frac{B_{r} B_{\phi}}{4\pi} \right) \equiv \int\mathrm{d} z \rho \alpha_{r \phi} c_{s}^{2}\equiv \Sigma \overline{\alpha_{r\phi}}c_{s}^{2} ,
    \label{eq:alpha_rp}
\end{equation}
where $c_{s}$ is isothermal sound speed on the midplane of the disk, and $\overline{\alpha_{r\phi}}$ is a non-dimensional parameter.

We consider the same limiting cases on the strength of the turbulence caused by magnetorotational instability (MRI) as in \citet{Suzuki+16}: MRI-active and MRI-inactive models. We take $\overline{\alpha_{r \phi}}=8\times 10^{-3}$ for the MRI-active model, and $\overline{\alpha_{r \phi}}=8\times 10^{-5}$ for the MRI-inactive model, respectively.

\subsection{Magnetic Disk Wind}
\label{DW}

Following \cite{Suzuki+16}, we use the $\alpha$-parameter prescription for the mass and angular momentum transport driven by the MHD disk wind, 
\begin{equation}
    \left( \rho \delta v_{\phi} v_{z} -\frac{B_{\phi}B_{z}}{4\pi} \right)_\mathrm{w}
    \equiv \left( \rho c_{s}^{2} \alpha_{\phi z}\right)_\mathrm{w}
    =\left( \rho c_{s}^{2} \right)_{\rm mid} \overline{\alpha_{\phi z}},
    \label{eq:alpha_pz}
\end{equation}
where $\overline{\alpha_{\phi z}}$ is a non-dimensional parameter, which is assumed to depend on the surface density $\Sigma (r,t)$ as
\begin{equation}
    \overline{\alpha_{\phi z}} (r,t) 
    = \mathrm{min} \left[10^{-5}\left( \frac{\Sigma}{\Sigma_{\rm int}}\right)^{-0.66},1 \right] ,
\end{equation}
where $\Sigma_{\rm int} (r)$ is the initial distribution of the surface density \citep{Bai+2013}.

We also assume that the mass loss rate by the MHD disk wind is proportional to $\left( \rho c_{s} \right)_{\rm mid}$ as 
\begin{equation}
    \left( \rho v_{z} \right)_{\mathrm{w}}=C_{\mathrm{w}} \left( \rho c_{s} \right)_{\rm mid},
    \label{eq:Cw}
\end{equation}
where $C_\mathrm{w}$ is a non-dimensional coefficient.
We impose a minimum or floor value to $C_\mathrm{w}$ by $C_\mathrm{w, 0}$. 
We take $C_\mathrm{w,0}=2\times 10^{-5}$ for the MRI-active model and $C_\mathrm{w,0}=1\times 10^{-5}$ for the MRI-inactive model, respectively.

\cite{Suzuki+16} provides the concrete formulae of $C_\mathrm{w}$ considering the energy balance to launch the disk wind. The energy equation is given by
\begin{multline}
    (\rho v_{z})_{\mathrm{w}}\left( E_{\mathrm{w}} + \frac{r^{2}\Omega^{2}}{2} \right) + F_{\mathrm{rad}} \\ = \frac{\Omega}{r} \left[ \frac{\partial}{\partial r} \left( r^{2} \Sigma \overline{\alpha_{r \phi}} c_{s}^{2} \right) +r^{2}\overline{\phi z} (\rho c_{s}^{2})_{\mathrm{mid}} \right] \\- \frac{1}{r} \frac{\partial}{\partial r} (r^{2} \Sigma \Omega \overline{\alpha_{r \phi}} c_{s}^{2}) ,
    \label{eq:COE}
\end{multline}
where $E_{\mathrm{w}}$ is the total specific energy of disk wind gas, which is assumed to be zero. 
On the left-hand side, the first and second terms represent the energy carried away by the disk wind and radiation, respectively. On the left-hand side, the first and second terms correspond to the energy released by accretion and that generated by viscous heating. 
In other words, Equation \eqref{eq:COE} describes how the energy available in the disk is distributed to the wind and radiation, which escape from the system. 
\cite{Suzuki+16} consider the following two extreme models of strong and weak disk wind (sDW and wDW). In wDW model, we assume that 10 $\%$ of the total available energy on the right-hand side of Equation \eqref{eq:COE} equals the energy carried away by the disk wind. The wDW model provides the formulae of $C_{\mathrm{w}}$ and $F_{\mathrm{rad}}$ as 
\begin{multline}
    C_{\mathrm{w}}=\mathrm{min}\Bigg[ C_{\mathrm{w, 0}}, \\ (1-\epsilon_{\mathrm{rad}})\left( \frac{3\sqrt{2\pi}c_{s}^{2}}{r^{2}\Omega^{2}}\overline{\alpha_{r\phi}} + \frac{2c_{s}}{r\Omega}\overline{\alpha_{\phi z}} \right) \Bigg] ,
    \label{eq:Cwe_WDW}
\end{multline}
\begin{equation}
    F_{\mathrm{rad}}=\epsilon_{\mathrm{rad}}\left( \frac{3}{2}\Omega \Sigma \overline{\alpha_{r\phi}}c_{s}^{2} + r\Omega \overline{\alpha_{\phi z}} \left( \rho c_{s}^{2} \right)_\mathrm{mid} \right) ,
    \label{eq:Frad_WDW}
\end{equation}
where $C_{\mathrm{w, 0}}$ is the upper limit of $C_{\mathrm{w}}$. 
In sDW model, on the other hand, we assume that the gravitational energy released by accretion equals the energy carried away by disk wind, leading to $C_{\mathrm{w}}$ and $F_{\mathrm{rad}}$ given by
\begin{multline}
    C_{\mathrm{w}}=\mathrm{min}\Bigg[ C_{\mathrm{w, 0}}, \\\mathrm{max}\left( \frac{2}{r^{3}\Omega \left(\rho c_{s}\right)_\mathrm{mid}} \frac{\partial}{\partial r}\left( r^{2} \Sigma \overline{\alpha_{r \phi}}C_{s}^{2} \right) \frac{2c_{s}}{r\Omega}\overline{\alpha_{\phi z}} , 0\right) \Bigg],
    \label{eq:Cwe_STDW}
\end{multline}
\begin{equation}
    F_{\mathrm{rad}}=\mathrm{max}\left( -\frac{1}{r}\left(r^{2}\Sigma \Omega \overline{\alpha_{r\phi}}c_{s}^{2} \right),0 \right) .
    \label{eq:Frad_STDW}
\end{equation}

With the above formulation, we rewrite equation~(\ref{eq:basic_eq1}) as 
\begin{multline}
    \frac{\partial \Sigma}{\partial t}=\frac{1}{r} \frac{\partial}{\partial r}\Bigg[ \frac{2}{r\Omega} \Bigg\{ \frac{\partial}{\partial r} \left( r^{2} \Sigma \overline{\alpha_{r\phi}}c_{s}^{2} \right) + r^{2}\left( \rho c_{s}^{2} \right)_{\rm mid} \overline{\alpha_{\phi z}} \Bigg\} \Bigg] \\ 
    -C_\mathrm{w} \left( \rho c_{s} \right)_{\rm mid}-\dot{\Sigma}_\mathrm{pw},
    \label{eq:basic_eq2}
\end{multline}
which is solvable with a given photoevaporation rate $\dot{\Sigma}_\mathrm{pw}$.

\subsection{Photoevaporation Model}
\label{photoeva}

We consider the photoevaporation caused by EUV and X-ray radiation coming from the central star, assuming that the total rate per unit area $\dot{\Sigma}_{\mathrm{pw}}(r)$ is the summation of EUV and X-ray driven photoevaporation rates,
\begin{equation}
   \dot{\Sigma}_{\mathrm{pw}} (r)= \dot{\Sigma}_{\mathrm{pw, EUV}} (r) + \dot{\Sigma}_{\mathrm{pw, X}} (r) .
\end{equation}
Following \citet{Nakatani+23}, we do not incorporate FUV photoevaporation in the present study, as we are primarily interested in small-grain-depleted disks, where FUV photoevaporation is ineffective. This population has the potential for the longest gas disk lifetimes; otherwise, PPDs are unlikely to survive beyond $10\Myr$ \citep{2015_Gorti, Komaki+21}. Focusing on the longest-living population is advantageous since it allows us to confidently rule out the primordial-origin scenario if the resulting lifetimes are found to be shorter than $\approx10\Myr$, even for this population. 
We discuss the plausibility of small-grain-depleted disks in Section~\ref{model_limit}.

The EUV photoevaporation rate $\dot{\Sigma}_{\mathrm{pw, EUV}} (r)$ is generally described as
\begin{equation}
    \dot{\Sigma}_{\mathrm{pw, EUV}} (r)=
    \begin{cases}
    2m_{\mathrm{H}} c_{s, \mathrm{H II}}(T_{\mathrm{H II}}) n_{0}(r) &\text{if } r\geq r_{\mathrm{crit}} ,\\
    0 &\text{if } r<r_{\mathrm{crit}}
    \end{cases}
    \label{eq:dSigma_photo_EUV_Tanaka}
\end{equation}
where $m_{\mathrm{H}}$ is mass of a hydrogen atom, $c_{s, \mathrm{H II}}$ the sound speed of the photoionized gas, and $n_{0} (r)$ the number density at the base of the photoionized layer \citep{Hollenbach1994}. 
\footnote{Strictly speaking, this assumption, along with the assumption of supersonic photoevaporative winds, holds only when the host star's luminosity exceeds a critical threshold \citep{Nakatani+24}. As a result, Equation \eqref{eq:dSigma_photo_EUV_Tanaka} somewhat overestimates the mass-loss rates when winds become gravity-inhibited for low luminosities. This introduces some uncertainties in the estimated disk lifetimes. Nevertheless, this does not alter our overall conclusions, and we believe that these uncertainties are minor compared to the larger uncertainties in disk parameters and the physics of disk evolution. }$r_{\mathrm{crit}}$ is the critical radius, and described $\beta r_{\mathrm{g}}$. $r_{\mathrm{g}}$ is the gravitational radius, and $\beta$ is 0.14 \citep{Weder+2023, Liffman_2003}.
For the base density $n_{0} (r)$, we use the formula presented by \cite{Tanaka+2013}, 
\begin{equation}
    n_{0}(r)=1.6 \times 10^{7} \left( \frac{\Phi_\mathrm{EUV}}{10^{49} \mathrm{s}^{-1}} \right)\left( \frac{r}{10^{15}\mathrm{cm}} \right)^{-3/2} \mathrm{cm^{-3}} ,
\end{equation}
where $\Phi_\mathrm{EUV}$ is the emissivity of the EUV photons. 
We assume that the temperature of photoionized gas ($T_{\mathrm{H II}}$) is $10^{4}$K. Note that our adopted $n_{0} (r)$ differs from that given by \cite{Hollenbach1994}, $n_0 \propto r^{-5/2}$, which has been widely used in previous studies including \cite{Nakatani+23} and \cite{Weder+2023}. While \cite{Hollenbach1994} derived $n_{0} (r)$ based on an idealized 1+1D modeling, \cite{Tanaka+2013} employed more elaborate modeling by solving radiation transfer in 2D to do that. 
In comparison to $\dot{\Sigma}_{\mathrm{pw, EUV}} (r)$ based on \cite{Hollenbach1994}, we consider the more efficient photoevaporation in the outer part of the disk.

For the X-ray photoevaporation rates $\dot{\Sigma}_{\mathrm{pw, X}} (r)$, we assume 10\% of the rates given by \cite{Owen+2012}. \cite{Sellek2024} have recently shown that \cite{Owen+2012} overestimated the rates due to their ignorance of some cooling processes involving oxygen atoms. We incorporate this update by applying the suppression factor of 0.1.
\footnote{
\cite{Sellek2024} also provide the surface mass-loss rates, but they became available during the preparation of this paper. Nevertheless, we expect using their fits would not change the conclusions.
}

If the MHD wind is efficiently launched from inner regions of the disk, the outer part can be shielded from stellar radiation. We consider such a shielding effect following \cite{Weder+2023}, where the significance of the shielding is modeled as a function of the column density of the disk wind from the central star to a given radius. We estimate the MHD wind column density as
\begin{equation}
    N_\mathrm{w}(r)=\int_{r_\mathrm{in}}^{r} \frac{\dot{\Sigma}_\mathrm{w}(r')}{2 \mu_{\mathrm{w}} m_\mathrm{H} v_{\mathrm{w}}} \mathrm{d} r' ,
    \label{eq:DWcolum_density}
\end{equation}
where $\dot{\Sigma}_\mathrm{w} = (\rho v_z)_\mathrm{w}$ is the local mass loss rate via the MHD disk wind, $\mu_{\mathrm{w}}$ is the mean molecular weight, $m_\mathrm{H}$ the hydrogen atom mass, and $v_{\mathrm{w}}$ the typical velocity of the disk wind. We take the specific values of $\mu_{\mathrm{w}}=2.23$ \citep{Weder+2023} and $v_{\mathrm{w}}=70\unit{km s^{-1}}$. We assume that at the radii where $N_\mathrm{w} > 10^{19}\mathrm{cm}^{-2}$ the EUV photoevaporation rate is zero. In addition, we also assume that X-ray photoevaporation is prevented in the further outer part of the disk where $N_\mathrm{w} > 10^{21}\mathrm{cm}^{-2}$, because X-ray heating occurs mainly in the layer whose column density is $\sim 10^{21}\mathrm{cm}^{-2}$ \citep{Alexander+2014}.

As in \cite{Nakatani+23}, we use the results of \cite{Kunitomo+21} to model the time evolution of the EUV and X-ray emissivity of the central star. 
We assume that the EUV emissivity is the sum of two components:
\begin{equation}
\label{eq:phiev}
\Phi_{\mathrm{EUV}} = \Phi_{\mathrm{EUV, ph}} + \Phi_{\mathrm{EUV, mag}},
\end{equation}
where $\Phi_{\mathrm{EUV, ph}}$ represents the thermal radiation from the stellar photosphere, and $\Phi_{\mathrm{EUV, mag}}$ the non-thermal emission caused by magnetic activity near the surface. Observations suggest an empirical relation between $\Phi_{\mathrm{EUV, mag}}$ and stellar X-ray luminosity $L_{\mathrm{X}}$:
\begin{equation}
    \log{\left(\frac{\Phi_{\mathrm{EUV, mag}}}{1~\mathrm{s}^{-1}}\right)}=20.40+0.66\log{\left( \frac{L_{\mathrm{X}}}{1~\mathrm{erg} \mathrm{s}^{-1}} \right)} ,
    \label{eq:Phi_EUVmag}
\end{equation}
with which we can estimate $\Phi_{\mathrm{EUV, mag}}$ from $L_{\mathrm{X}}$. \citet{Kunitomo+21} derive the X-ray luminosity $L_{\mathrm{X}}$ from stellar evolution calculations using the relation 
\begin{equation}
    L_{\mathrm{X}}=\mathrm{min}(10^{-3.13}, 5.3\times 10^{-6} Ro^{-2.7})L_{\mathrm{*}} ,
    \label{eq:Lx}
\end{equation}
where $L_{\mathrm{*}}$ is the stellar 
bolometric
luminosity, $Ro \equiv P_{\mathrm{rot}} / \tau_{\mathrm{conv}}$ the Rossby number, $P_{\mathrm{rot}}$ the rotational period, and $\tau_{\mathrm{conv}}$ the convective turnover timescale \citep[see also][]{Wright2011}. 
\cite{Kunitomo+21} assume a fixed value of $P_{\mathrm{rot}} = 3 \unit{days}$.
 Following \cite{Nakatani+23}, we do not impose a minimum X-ray luminosity, unlike \cite{Kunitomo+21}. Note that the minimum stellar X-ray luminosity is due to the sensitivity limit of the observations.

Equation~\eqref{eq:Lx} shows that the shorter turnover timescale $\tau_{\mathrm{conv}}$ leads to the lower X-ray luminosity. 
Since the EUV emissivity depends on the X-ray luminosity through Equation~\eqref{eq:Phi_EUVmag}, both of these quantities vary with $\tau_{\mathrm{conv}}$, i.e., how large a surface convective layer appears. This means that the evolution of the interior structure of pre-main-sequence stars determines both the EUV and X-ray photoevaporation rates, and thus affects the evolution of protoplanetary disks.

\begin{figure}
    \centering
    \includegraphics[width=8cm]{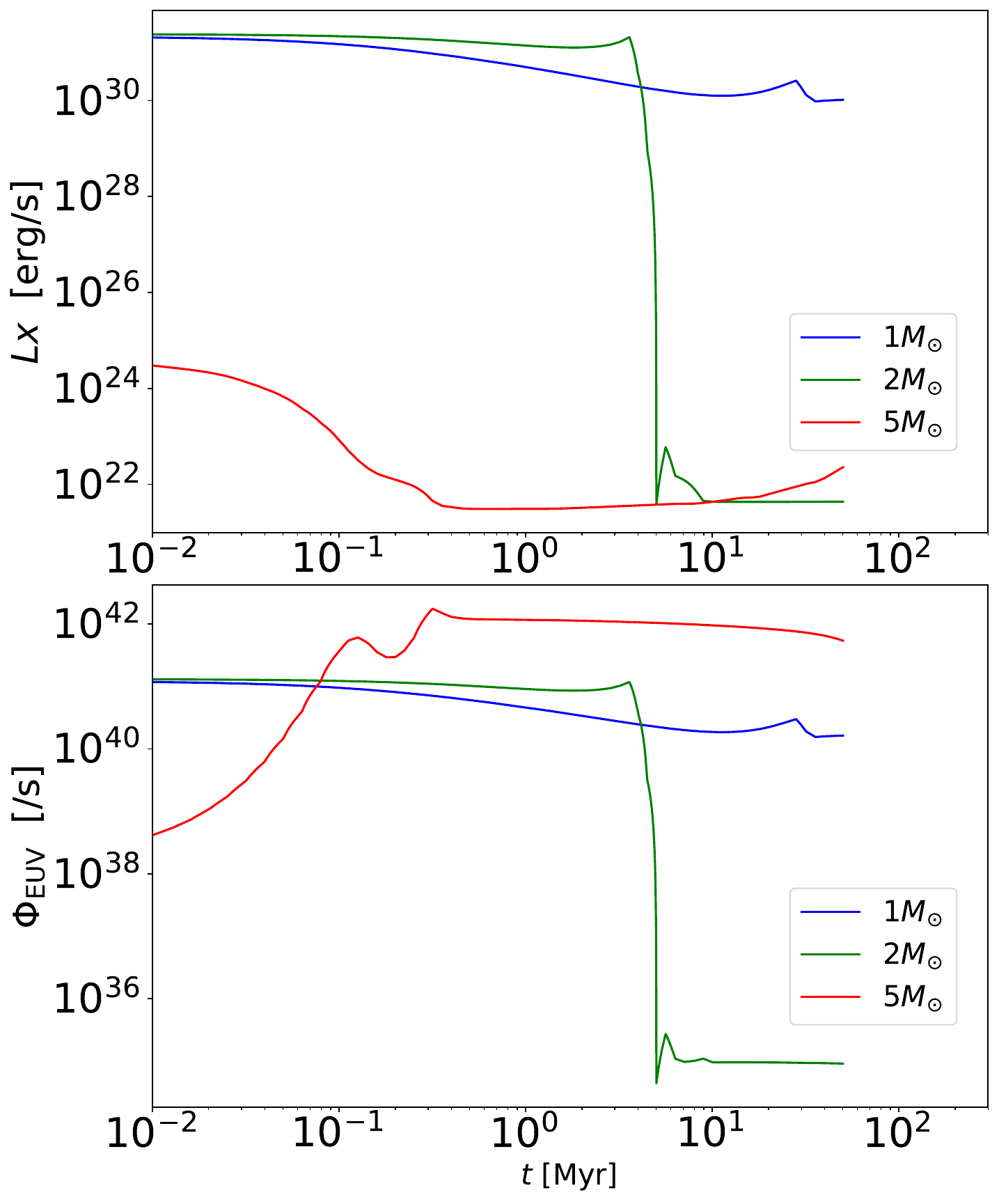}
    \caption{Time evolution of stellar X-ray luminosity and EUV photon emission rates of
    with different masses. 
    The upper and lower panels show the evolution of the X-ray luminosity and EUV emissivity as functions of the age, respectively. In each panel, the blue, green, and red lines represent the star of $1~\Msun$, $2~\Msun$, and $5~\Msun$, respectively.
    }
    \label{fig:Lx_EUV}
\end{figure}

Figure~\ref{fig:Lx_EUV} shows the time evolution of the X-ray luminosity $L_\mathrm{X}$ and EUV photon emissivity $\Phi_\mathrm{EUV}$ for different central stars with $M_{*}=1$, 2, and $5\Msun$.
In the model with $M_{*}=1\Msun$, both the X-ray luminosity and EUV emissivity remain high until the age of $\gtrsim 10\unit{Myr}$. In the model with $M_{*}=2\Msun$, on the contrary, both $L_\mathrm{X}$ and $\Phi_\mathrm{EUV}$ dramatically decrease at age $\simeq 4\unit{Myr}$. 
In the model with $M_{*}=5\Msun$, $\Phi_\mathrm{EUV}$ is initially small and then increases to become the largest after $\sim 0.1\unit{Myr}$, while $L_\mathrm{X}$ is small throughout the evolution.

The variations described above can be understood in terms of stellar evolution. 
In the model of $M_{*}=1 \Msun$, the surface convective layer persists throughout the evolution. However, in the model of $M_{*}=2 \Msun$ it disappears in the middle of the evolution, corresponding to the sharp decrease in $L_\mathrm{X}$ and $\Phi_\mathrm{EUV}$.
The photospheric EUV component $\Phi_\mathrm{EUV,ph}$ has nothing to do with the presence of the surface convective layer, but its contribution is also small because of the low effective temperature. In the model with $M_{*}=5 \Msun$, the surface convective zone is small from the beginning, leading to the initial low values of $L_\mathrm{X}$ and $\Phi_\mathrm{EUV}$. Only $\Phi_\mathrm{EUV}$ increases significantly later because the effective temperature and thus $\Phi_\mathrm{EUV,ph}$ rise through the pre-main-sequence evolution for this model.

In addition to the two components of $\Phi_{\mathrm{EUV, ph}}$ and $\Phi_{\mathrm{EUV, mag}}$ in Equation \eqref{eq:phiev}, mass accretion on the central star can also contribute to the EUV emissivity, as a part of the released gravitational energy can be radiated away. However, we ignore this additional component $\Phi_{\mathrm{EUV, acc}}$ in Section~\ref{result} for simplicity.
Whereas \cite{Nakatani+23} assumes that 4\% of the gravitational energy is converted to EUV radiation energy, the energy conservation rate is highly uncertain. We separately study the effect of $\Phi_{\mathrm{EUV, acc}}$ in the appendix.

\subsection{Temperature in Disk Mid-plane}
\label{Temp}

The remaining unknown quantity to integrate Equation \eqref{eq:basic_eq2} is $c_s$, or the temperature in the mid-plane of the disk $T$. Following \citet{Suzuki+16}, we use a simple model by \citet{Nakamoto+1994}, where $T$ is written as 
\begin{equation}
    T^{4}=T_{\mathrm{irr}}^{4}+T_{\mathrm{vis}}^{4} ,
    \label{eq:T_eq}
\end{equation}
where $T_{\mathrm{irr}}$ is the equilibrium temperature resulting from the balance between irradiation heating by the central star and dust cooling, and $T_{\mathrm{vis}}$ represents the equilibrium between viscous heating and dust cooling.
The former temperature $T_{\mathrm{irr}}$ is given by
\begin{equation}
    T_{\mathrm{irr}}=T_{1\mathrm{au}}\left( \frac{r}{1\mathrm{au}} \right)^{-1/2} \left( \frac{L_{\mathrm{*}}}{L_{\odot}} \right)^{1/4} ,
    \label{eq:T_irr}
\end{equation} 
where $T_{1\mathrm{au}}$ represents $T_\mathrm{irr}$ at the radius of 1~$\unit{au}$ around the Sun, which we estimate $T_{1\mathrm{au}} = 280$~K.
The latter temperature $T_{\mathrm{vis}}$ is obtained by solving
\begin{equation}
    2\sigma_{\mathrm{SB}}T_{\mathrm{vis}}^{4}=\left( \frac{3}{8} \tau_{\mathrm{R}} + \frac{1}{2\tau_{\mathrm{P}}} \right) F_{\mathrm{rad}} ,
    \label{eq:T_vis}
\end{equation}
where $\tau_{\mathrm{R}} = 0.5 \times \kappa_{\mathrm{R}}\Sigma$ is the optical depth measured at the midplane defined with the Rosseland opacity $\kappa_{\mathrm{R}}$, and $\tau_{\mathrm{P}}$ is the same but for the Planck mean opacity $\kappa_{\mathrm{P}}$. 
We approximate the Planck mean optical depth as $\tau_{\mathrm{P}}=\mathrm{max}(2.4\tau_{\mathrm{R}}, 0.5)$, for which we impose the lower bound to ensure that the pre-factor in Equation \eqref{eq:T_vis} converges as $\left( \frac{3}{8} \tau_{\mathrm{R}} + \frac{1}{2\tau_{\mathrm{P}}} \right) \rightarrow 1 $ at the optically thin limit.
For the dust opacity $\kappa_{\mathrm{R}}$, we use the same formula as in \cite{Kunitomo2020} and \cite{Komaki+2023}.

Note that our model assumes dust opacity with the standard dust-to-gas mass ratio and grain size distribution, which appears to be inconsistent with our assumption of very small-grain depletion. However, if the very small grains were depleted only from the disk’s atmosphere where photoevaporation occurs, and the micron-sized grains governing $T_{\rm vis}$ were present only below the disk's photosphere, using standard dust opacity does not contradict the assumption of very small-grain depletion. In addition, since opacity only weakly affects $T_{\rm vis}$, dust opacity hardly impacts disk evolution. We also stress that $T_{\rm vis}$ exceeds $T_{\rm irr}$ only at specific radii and predominantly during the early stages in most of our simulations. We have confirmed these points by performing tests where we reduce the opacity in our fiducial model by an order of magnitude; the result remains almost the same.

\subsection{Initial and Boundary Conditions}
\label{initial_condi}

The mass of the disk in the initial state $M_\mathrm{disk,0}$ is an input parameter in our calculations. The initial distribution of the gas surface density is assumed as 
\begin{equation}
    \Sigma_{\mathrm{int}}=\Sigma_{1\mathrm{au}} \left(\frac{r}{1\mathrm{au}}\right)^{-3/2}\exp{\left(-\frac{r}{r_{\mathrm{cut}}}\right)} ,
    \label{eq:sgm_ini}
\end{equation}
where $r_\mathrm{cut}$ is the outer cut-off radius, for which we adopt $r_{\mathrm{cut}}=30$~$\unit{au}$ as the default value. We determine the value of $\Sigma_{1\mathrm{au}}$ to make the integrated total gas mass equal to $M_\mathrm{disk,0}$.

We follow the evolution of $\Sigma(r,t)$ by solving Equation~(\ref{eq:basic_eq2}) in the radial area between $r_{\mathrm{in}} =0.01$~au and $r_{\mathrm{out}} =10^4$~au.
We divide the radial computational domain into 3000 numerous cells to differentiate Equation~(\ref{eq:basic_eq2}). We set the width of a cell $\delta r$ to increase in proportion to $r^{1/2}$. 
We impose the following boundary condition at $r_{\mathrm{in}}$ and $r_{\mathrm{out}}$
\begin{equation}
    \frac{\partial}{\partial r} (\Sigma r)=0 ,
\end{equation}
which is consistent with the zero-torque condition of \cite{Linden-Bell+1974} \citep{Suzuki+22}.

\subsection{Models Considered}
\label{disk_model}

Table~\ref{tab:model} summarizes our models considered. In the fiducial (FID) models, we use the weak DW and MRI inactive models. We assume that $M_\mathrm{disk,0}$ is proportional to the mass of the central star $M_*$ with a fixed ratio $M_\mathrm{disk,0}/M_* = 0.1$. We ignore the EUV radiation generated by mass accretion and include the shielding effect. 
We additionally investigate models with strong DW models (models sDW), MRI active models (MRIact), reduced initial disk mass (Mdisk0.03 and Mdisk0.01), including EUV radiation generated by mass accretion (EUVACC), and ignoring the shielding effect (NoShield). To isolate the influence of varying each input parameter, we keep the remaining parameters unchanged for these models.

We update the surface density $\Sigma(r,t)$ by integrating a discretized form of Equation \eqref{eq:basic_eq2} using an implicit method for all models. We define the lifetime of the disk as the epoch when the mass of the disk $M_\mathrm{disk}$ falls to nearly zero. 
Since the decline of $M_\mathrm{disk}$ is very rapid in the latest stage for all the models, our derived disk lifetime hardly changes if defined as when $M_\mathrm{disk} < 10^{-4} M_\mathrm{disk,0}$. 
In observations, disk lifetimes are typically estimated based on infrared (IR) observations of the inner disk, rather than the entire disk. We estimate the lifetime of the inner disk in Appendix~\ref{sec:IR_lifetime}.
Given our focus on the primordial origins of gas-rich debris disks, we do not evaluate the lifetime solely with the 
dispersal of an inner part ($\lesssim 10\unit{au}$), whereas near-infrared radiation from it often marks the presence of the disk.

\begin{table*}
    \centering
    \caption{models considered}
    \centering
    \movetableright=-1.1in
    \begin{tabular}{c| c c c c c c c}
        
        Name of the model & $\Phi_{\mathrm{EUV, acc}}$ & shielding & disk wind & $M_{\mathrm{disk, 0}}/M_{*}$ & MRI & $r_{\mathrm{cut}}$ [au] & Section\\
        
        \hline
        FID-N & No & Yes & Weak & 0.1 & Inactive & 30 & \ref{fiducial}\\

        NoShield-N & No & No & Weak & 0.1& Inactive & 30 & \ref{shield}\\
        
        sDW-N & No & Yes & Strong & 0.1& Inactive & 30 & \ref{AM}\\
        
        MRIact-N & No & Yes & Weak & 0.1& active & 30 & \ref{AM}\\
        
        Mdisk0.03-N & No & Yes & Weak & 0.03& Inactive & 30 & \ref{AM}\\
        
        Mdisk0.01-N & No & Yes & Weak & 0.01& Inactive & 30 & \ref{initial_diskmass}\\
        
        Rcut1-N & No & Yes & Weak & 0.1& Inactive & 30 $\times$ N & \ref{initial_diskmass}\\
        
        EUVACC-N & Yes & Yes & Weak & 0.1& Inactive & 30 & Appendix~\ref{sec:accretion-generated_EUV}\\
            
    \end{tabular}
    \tablecomments{The suffix $N$ at the end of each model name, which takes values of 1, 1.5, 2, 3, 4, and 5, denotes the mass of the central star in the solar mass unit, while the other parts refer to the model parameters in each row. For instance, FID-2 represents the fiducial run with a $2~\Msun$ host star. The MRI active and inactive models assume $\overline{\alpha_{r \phi}} = 8\times 10^{-3}$ and $10^{-5}$, respectively.}
    \label{tab:model}
\end{table*}

\section{results}
\label{result}
In this section, we present the derived lifetimes as a function of stellar mass and discuss how they vary depending on the model setups. 
In Section~\ref{fiducial}, we show the results of the fiducial (FID) models.
We also study the effects of varying the input physics in the following sections; shielding of stellar radiation by the MHD disk wind in Section~\ref{shield}, varying the efficiency of angular momentum transport through the disk in Section~\ref{AM}, and the dependence on the initial disk parameters in Section~\ref{initial_diskmass}.

\subsection{Fiducial Model}
\label{fiducial}

\subsubsection{Stellar-mass Dependence of Evolution}
\label{results_fiducial} 
\begin{figure}
    \centering
    \includegraphics[width=8cm]{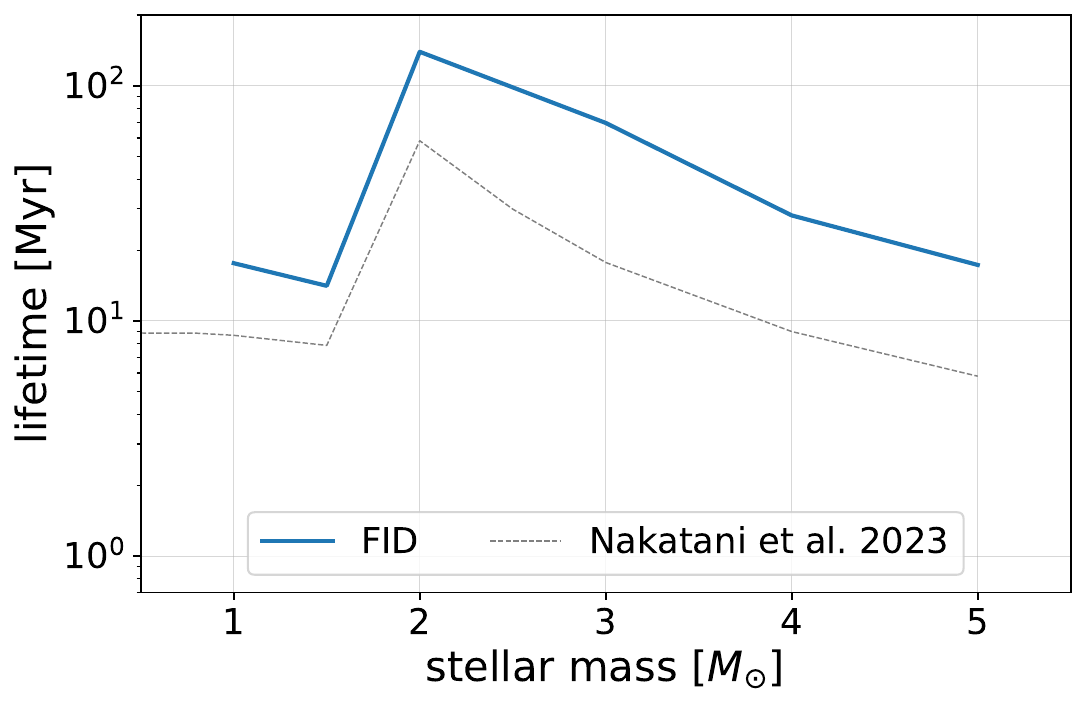}
    \caption{
Dependence of protoplanetary disk lifetimes on stellar mass. The blue line illustrates our 1D disk models using standard input physics (models FID in Table~\ref{tab:model}). The thin dotted line represents the one-zone models of \cite{Nakatani+23}, for which only stellar EUV radiation is considered for photoevaporation.
    }
    \label{fig:lifetime_fiducial}
\end{figure}
The green solid line of Figure~\ref{fig:lifetime_fiducial} shows the dependence of disk lifetime on stellar mass in the fiducial (FID) models. Among these, the lifetime of the disk has a maximum at $M_* = 2\Msun$, which is consistent with the current observational trend that gas-rich debris disks are frequently detected around early A~stars \citep[e.g.,][]{Moor2017}. 
At $M_* = 2\Msun$, the disk lifetime can even exceed $100 \Myr$, indicating the potential for significant longevity in initially massive, small-grain-depleted disks around early A~stars. 
We note, however, that this result should be interpreted within the application limit of our model (Section~\ref{model_limit}).

\begin{figure}
    \centering
    \includegraphics[width=8cm]{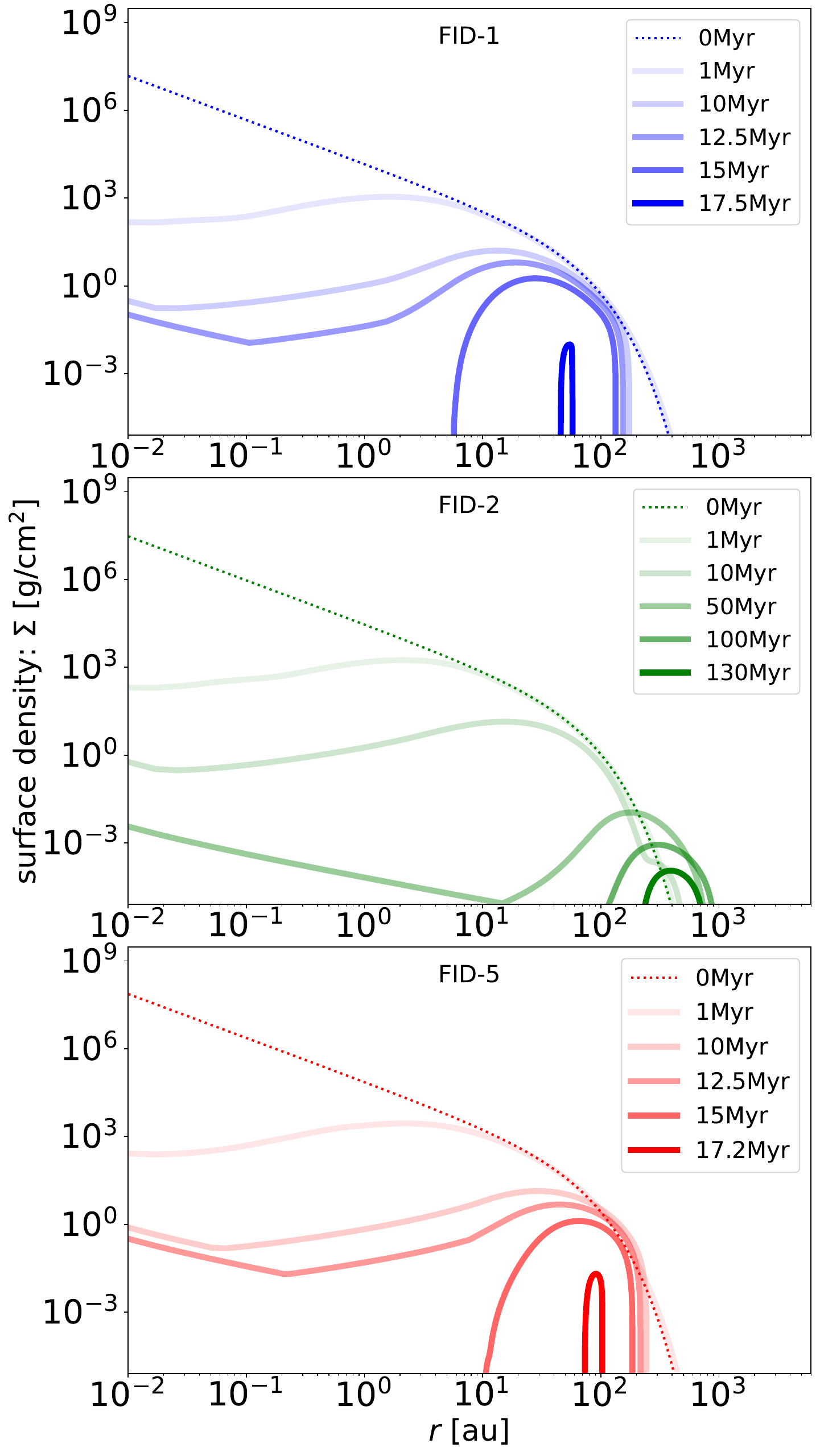}
    \caption{Evolution of the gas surface density as functions of the distance from the host star. In each panel, the different lines correspond to the different epochs, for which the darker color represents the later snapshot.    
    The top, middle, and bottom panels represent different masses of the host star, $1\Msun$, $2\Msun$, and $5\Msun$ in descending order (i.e., models FID-1, FID-2, and FID-5 in Table~\ref{tab:model}). 
    }
    \label{fig:fiducial_Sigma}
\end{figure}

The lifetime peak at $M_* = 2 \Msun$ (Figure~\ref{fig:lifetime_fiducial}) was also predicted by the one-zone model of \cite{Nakatani+23}, although their disk lifetimes are systematically shorter than those of our model. This discrepancy arises from the shielding effect and the overestimation of photoevaporation rates in \citet{Nakatani+23}, where they use a radially integrated photoevaporation rate $\dot{M}_\mathrm{ph}$, without considering the radial distribution of disk gas. This approach assumes that mass is removed at the same rate regardless of whether the disk has a continuous or ring-like configuration (cf. Figure~\ref{fig:fiducial_Sigma}), which overlooks the fact that photoevaporation cannot occur in regions where gas has already 
dispersed.

Figure~\ref{fig:fiducial_Sigma} shows the evolution of the surface density for different host stars with masses $M_* = 1$, 2, and $5\Msun$ (models FID-1, 2, and 5).
In models FID-1 and FID-5, an outer disk ($> 200 \unit{au}$) first 
disperses, and then an inner disk ($< 10\unit{au}$) gradually disappears. The remaining mass in $10 - 100\unit{au}$  
disperses finally, which is no later than $2\unit{Myr}$. Conversely, in model FID-2, the mass at $> 100 \unit{au}$ persists for over $10\unit{Myr}$.
The debris disks usually have dust rings at $\sim 10 - 100\unit{au}$ \citep{Hughes_2018}, which are also the typical radii the gas components cover for the gas-rich debris disks. 
This trend aligns with our FID-2 showing that the gas survives long around $10 - 100\unit{au}$.

\begin{figure}
\centering
\includegraphics[width=8cm]{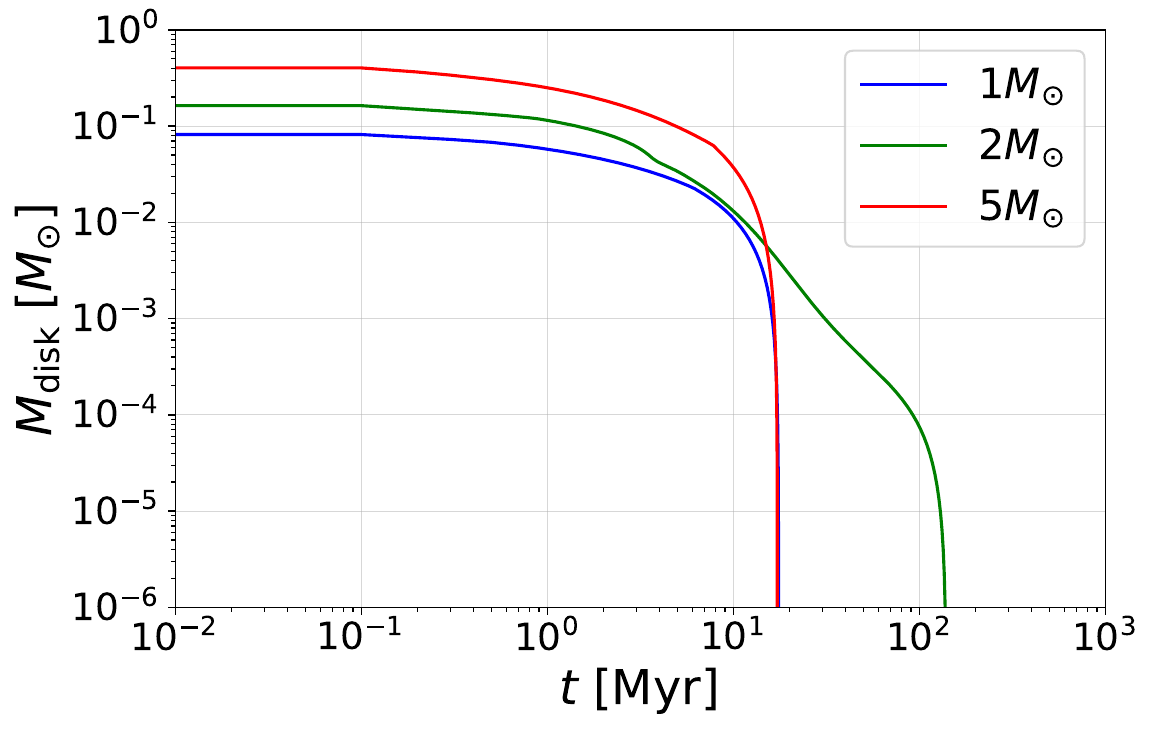}
\caption{Time evolution of the disk mass for the fiducial (FID) models. The blue, green, and red lines represent the models with different central stars with $M_* = 1$, 2, and $5\Msun$ (i.e., FID-1, 2, and 5), respectively.
}
\label{fig:fiducial_M_disk}
\end{figure}

Figure~\ref{fig:fiducial_M_disk} shows the time evolution of the disk mass $M_\mathrm{disk}$ for the fiducial models, FID-1, 2, and 5. In FID-1 and 5, the disk mass does not change much until $t \gtrsim 10\unit{Myr}$, after which it rapidly decreases around $t \simeq 20\unit{Myr}$. This rapid decrease in $M_\mathrm{disk}$ signifies a shift in the dominant dispersal mechanism from mass accretion to photoevaporation. In contrast, in FID-2, $M_\mathrm{disk}$ gradually decreases over an extended period of $10\unit{Myr} \lesssim t \lesssim 100\unit{Myr}$. For a host star with $2\Msun$, the EUV and X-ray photoevaporation rates drop significantly for $t \gtrsim 4\unit{Myr}$ due to the disappearance of the surface convective layer (Figure~\ref{fig:Lx_EUV}). Hence, effective disk dispersal through photoevaporation does not start until $M_\mathrm{disk}$ decreases sufficiently. We return to this point also in Section~\ref{timescale_fiducial}.

\subsubsection{Comparisons of Timescales}
\label{timescale_fiducial}

\begin{figure}
    \centering
    \includegraphics[width=8cm]{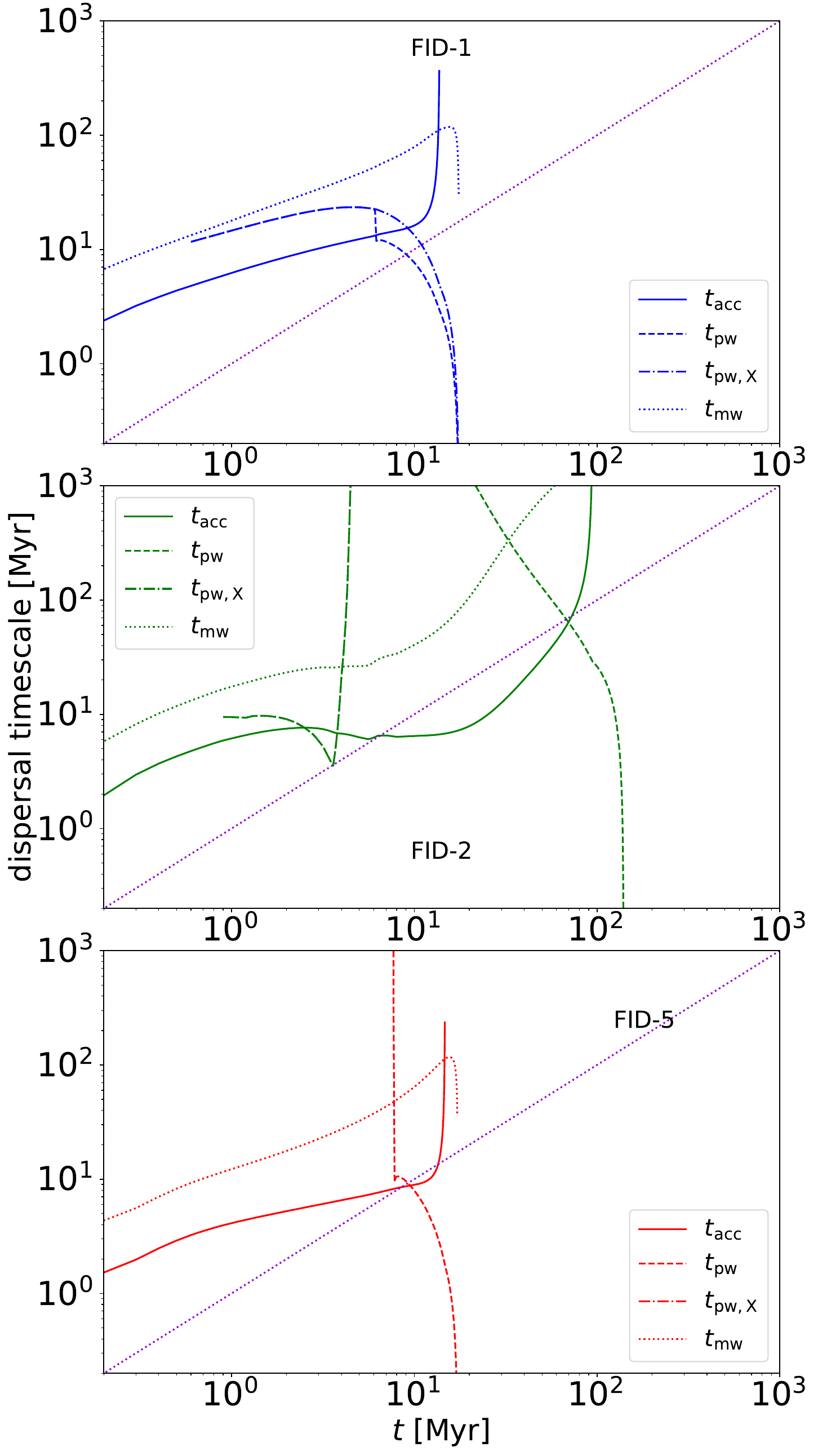}
    \caption{
Evolution of various timescales related to the disk dispersal. The top, middle, and bottom panels represent the fiducial models with the central star with $1$, 2, and $5\Msun$ (FID-1, 2, and 5 in Table~\ref{tab:model}). In each panel, the solid, dashed, dot-dashed, and dotted lines represent the timescales of the mass accretion $t_\mathrm{acc}$, photoevaporation by EUV and X-ray radiation $t_\mathrm{pw}$, photoevaporation only by X-ray radiation $t_\mathrm{pw, X}$, and MHD disk wind (excluding accretion by angular momentum transport by MHD disk wind) $t_\mathrm{DW}$. As shown by Equations \eqref{eq:tacc} - \eqref{eq:tdw}, these are defined as the disk mass divided by the corresponding dispersal rates. The diagonal violet dotted line represents the elapsed time, indicating when each process becomes effective. 
    }
    \label{fig:fiducial_t_dispersal}
\end{figure}

In this section, we examine the processes that determine the gas disk lifetimes in our fiducial models. 
We define the timescales for dispersal due to viscous accretion, EUV plus X-ray photoevaporation, X-ray photoevaporation alone, and mass loss via MHD winds as follows:
\begin{equation}
\label{eq:tacc}
    t_{\mathrm{acc}}(t)=\frac{M_{\mathrm{disk}}(t)}{\dot{M}_{\mathrm{acc}}(t)} ,
\end{equation}
\begin{equation}
    t_{\mathrm{pw}}(t)=\frac{M_{\mathrm{disk}}(t)}{\dot{M}_{\mathrm{pw}}(t)} ,
\end{equation}
\begin{equation}
    t_{\mathrm{pw, X}}(t)=\frac{M_{\mathrm{disk}}(t)}{\dot{M}_{\mathrm{pw,X}}(t)} ,
\end{equation}
\begin{equation}
\label{eq:tdw}
    t_{\mathrm{mw}}(t)=\frac{M_{\mathrm{disk}}(t)}{\dot{M}_{\mathrm{mw}}(t)}, 
\end{equation}
where $\dot{M}_\mathrm{acc}$ is the mass accretion rate onto the central star, and the other $\dot{M}$ are the total mass-loss rates integrated over the whole disk. The suffixes of "mw" and "pw" refer to the magnetic disk wind and the photoevaporation wind, respectively.
These timescales are particularly useful because the shortest timescale indicates the dominant dispersal process at a given moment. Additionally, the disk lifetime corresponds to the point where the timescale for the dominant process falls below the system's age.

In FID-1 (top panel of Figure~\ref{fig:fiducial_t_dispersal}), accretion initially dominates the dispersal process. However, as the shielding becomes ineffective around $5\unit{\Myr}$, photoevaporation takes over as the primary dispersal process. The corresponding dispersal timescale $t_\mathrm{pw}$ approaches zero as the disk mass decreases. The contribution of MHD disk winds is negligible in this model, meaning that the evolution essentially follows the UV-switch model by \citet{2001_Clarke}.

In FID-2 model (middle panel of Figure~\ref{fig:fiducial_t_dispersal}), $t_{\mathrm{pw}}$ increases rapidly at $4 \unit{Myr}$ because the X-ray luminosity ($L_{\rm X}$) and the EUV emissivity of the surface magnetic activity ($\Phi_{\mathrm{EUV, mag}}$) drop dramatically due to the disappearance of the surface convective zone. 
Until $\simeq 60 \unit{\Myr}$, $t_{\mathrm{acc}}$ is the shortest timescale and less than age, 
signifying a gradual reduction in disk mass through accretion. This period aligns with the slow decrease phase of $M_\mathrm{disk}$ after $\sim 10\unit{\Myr}$ illustrated in Figure \ref{fig:fiducial_M_disk}.
Despite the small values of $L_{\rm X}$ and $\Phi_{\mathrm{EUV, mag}}$ during this period (Figure \ref{fig:Lx_EUV}), the photoevaporation timescale $t_{\mathrm{pw}}$ gradually decreases as $M_\mathrm{disk}$ decreases.
At the epoch of $\simeq 60 \unit{Myr}$, $t_{\mathrm{pw}}$ becomes the shortest, and the disk is  
dispersed by $t_{\mathrm{pw}}\rightarrow 0$. 
This shows that in this model mass accretion  
disperses the disk first, and then photoevaporation dominates after the surface density decreases small enough.
The disappearance of the convective zone makes the photoevaporation rate small, and the lifetime of the disk becomes longer than $1 \Msun$. The contribution of mass loss by MHD disk wind is negligible as in FID-1.

In the model of FID-5 (bottom panel of Figure~\ref{fig:fiducial_t_dispersal}), $t_{\mathrm{pw}}$ remains above $1000 \unit{\Myr}$ up to approximately $20 \unit{Myr}$. This is because the surface convective layer is absent from early on, leading to minimal contributions from $L_{\rm X}$ and $\Phi_{\mathrm{EUV, mag}}$. Nevertheless, due to the high effective stellar temperature, $\Phi_{\mathrm{EUV, ph}}$ increases (Figure~\ref{fig:Lx_EUV}), and $t_{\mathrm{pw}}$ dramatically decreases around $\simeq 20 \unit{Myr}$ as the shielding effect of EUV radiation ceases. Consequently, $t_{\mathrm{pw}}$ falls below both $t_{\mathrm{acc}}$ and the age, which accelerates the disk dispersal through photoevaporation. As in FID-1 and FID-2, the impact of mass loss by MHD disk wind is negligible, as $t_{\mathrm{mw}}$ never becomes the shortest throughout the evolution.

\subsection{Effects of Shielding}
\label{shield}

\begin{figure}
    \centering
    \includegraphics[width=8cm]{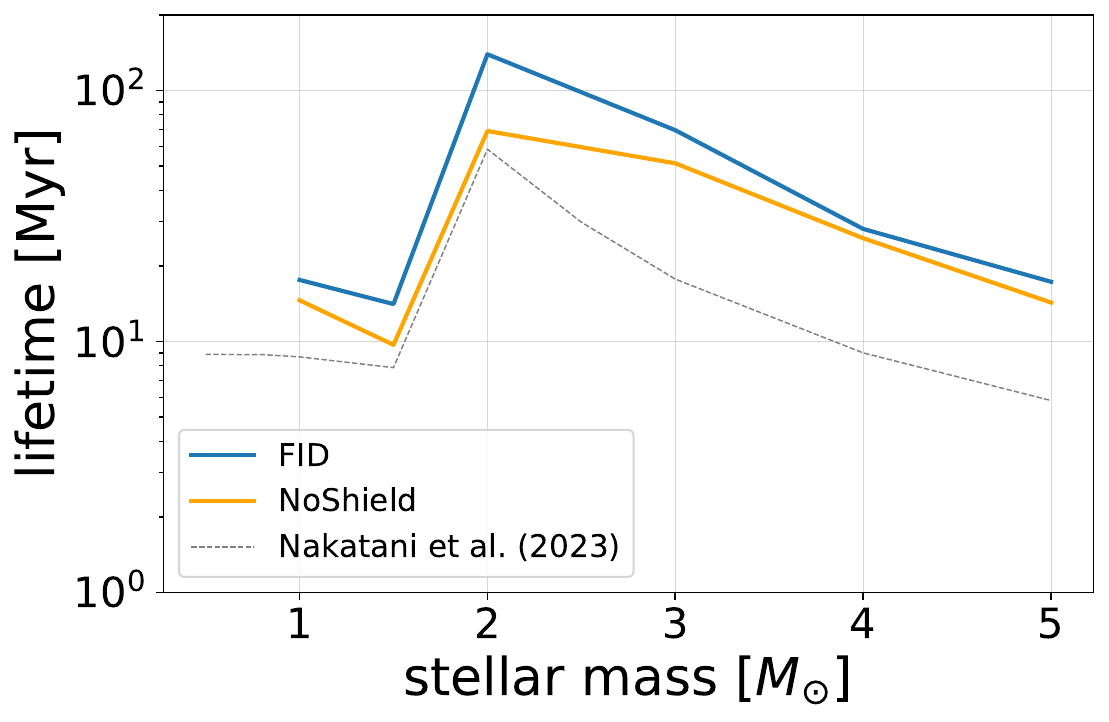}
    \caption{
Effect of shielding the stellar radiation by MHD disk wind on the disk lifetime. The blue and thin dotted lines show the stellar-mass dependence of the disk lifetime of our fiducial models and one-zone models by \cite{Nakatani+23}, respectively. The orange line represents NoShield models, where the shielding effect is ineffective (see Table~\ref{tab:model}).
}
\label{fig:lifetime_Noshield}
\end{figure}

Figure~\ref{fig:lifetime_Noshield} compares the stellar mass dependence of disk lifetime between FID and NoShield models (Table~\ref{tab:model}). 
We see that, without the shielding effect of the stellar radiation by MHD disk wind, the lifetime becomes slightly shorter than with the shielding effect. This is reasonable because the photoevaporation starts to contribute to the disk dispersal early on in NoShield models. However, the overall result that the lifetime takes the maximum around $2\Msun$ does not change.

The figure further shows that NoShield models provide more consistent results with the one-zone models than FID models. This suggests that the shielding effect also contributes to explaining the discrepancy between our 1D and previous one-zone models, because it is only considered in the 1D models. As already mentioned in Section \ref{results_fiducial}, the overestimation of the photoevaporation rates in one-zone models is another contributing factor.

\subsection{Enhancing MRI Viscosity and MHD Disk Wind}
\label{AM}
\begin{figure}
    \centering
    \includegraphics[width=8cm]{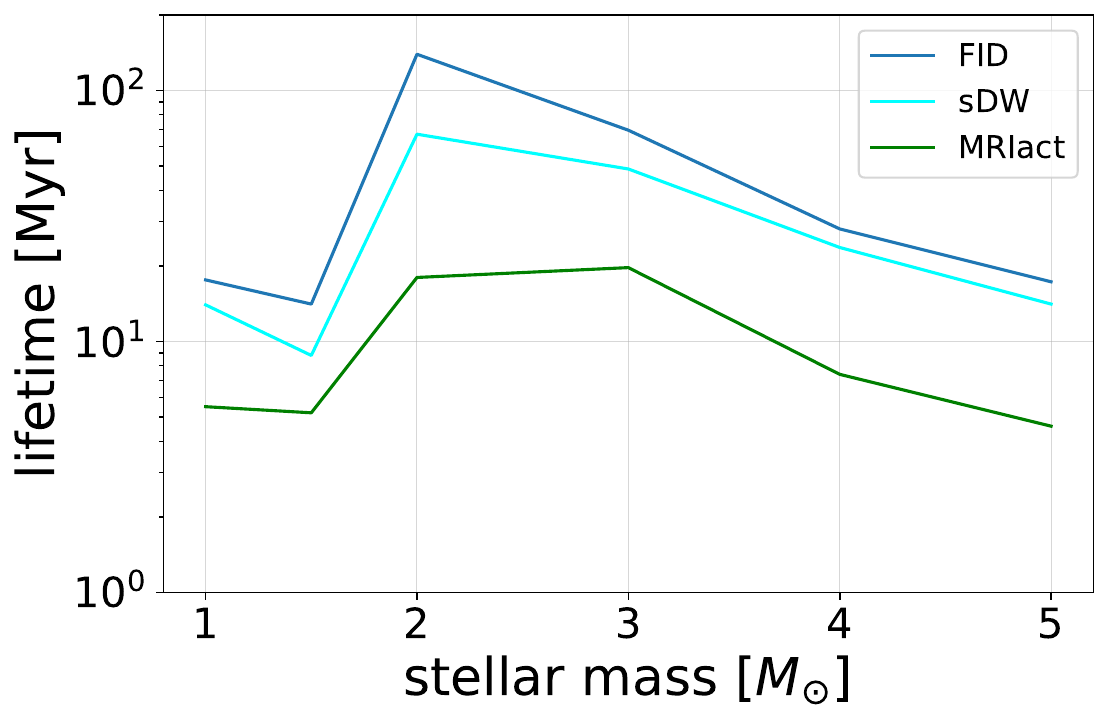}
    \caption{
Effect of varying MHD disk wind models on the disk lifetime. The blue line shows the stellar-mass dependence of the disk lifetime of our fiducial models. The cyan and green lines represent sDW and MRIact models, respectively (see Table~\ref{tab:model}).
}
    \label{fig:lifetime_all_AM}
\end{figure}

Figure~\ref{fig:lifetime_all_AM} shows the stellar mass dependence of the disk lifetime with different DW models, comparing the results of FID, MRIact, and sDW models (Table \ref{tab:model}). The MRIact models incorporate scenarios where MRI is active. In sDW models, we apply the strong DW model of \cite{Suzuki+16}. 
We find that the lifetimes in the MRIact and sDW models are shorter than those in FID models. Nevertheless, the stellar-mass dependence is overall similar, peaking at $M_* = 2$--$3 \Msun$. 
The MRIact models result in the shortest lifetimes, and they exceed $10\Myr$ only slightly even at $M_*= 2$--$3 \Msun$. 
This suggests that sufficiently small $\alpha$ values ($\leq10^{-2}$) are necessary for protoplanetary disks to evolve into gas-rich debris disks. 

There are several observational estimates for $\alpha$ in PPDs, and generally, small values are preferred.  
For example, observations of CO(3-2) in HD~163296 suggest $\alpha < 10^{-3}$ \citep{Flaherty_2015}. 
\cite{Dullemond_2018} suggest that low-viscosity disks better explain the observed ring structures.
This indicates that disk evolution scenarios with a large $\alpha$, as seen in our MRIact models, may be relatively rare.
Note that the preference for small $\alpha$ contrasts with some secondary-origin models \citep[$\alpha \sim 0.1$--$1.5$][]{Kral2016, Marino2020}.

\begin{figure}
    \centering
    \includegraphics[width=8cm]{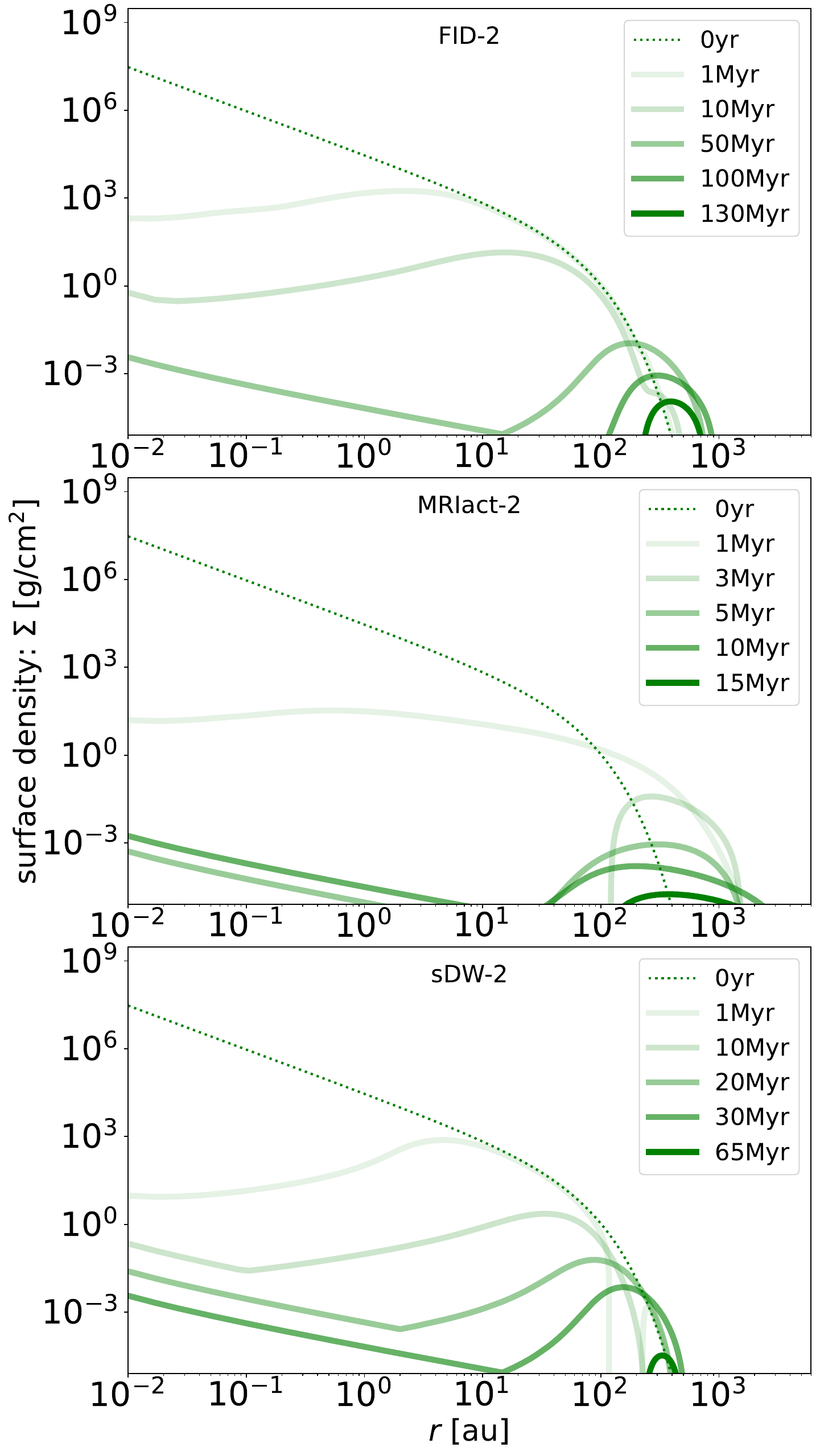}
    \caption{Evolution of the surface density around the central star with $2\Msun$ with different MRI viscosity and MHD disk wind models. The top, middle, and bottom panels represent models FID-2, MRIact-2, and sDW-2, respectively (see Table~\ref{tab:model} for model details).
    }
    \label{fig:Sigma_AM}
\end{figure}

Figure~\ref{fig:Sigma_AM} shows the evolution of the gas surface density for FID-2, MRIact-2, and sDW-2 models. In MRIact-2 and sDW-2 models, the surface density $\Sigma$ decreases more quickly compared to FID-2 model. MRIact-2 model shows that the disk gas extends farther out than FID-2, with the outer edge exceeding $\sim 10^3\unit{au}$ after $\sim 1\unit{\Myr}$. This is attributed to the increased angular momentum transport promoting efficient gas extension. In addition, an inner part ($r \lesssim 100\unit{au}$) of the disk temporarily 
disperses around $3\Myr$. Afterward, the inner part is refilled by the gas supply from the outer part and reemerges for the period of $5 \Myr \lesssim t \lesssim 10 \Myr$.
The inner disk 
disperses again by the epoch of $\simeq 15 \Myr$, and finally the remaining outer disk 
disperses soon after that. 
In the sDW-2 model, a significant rapid decline in $\Sigma$ is observed after $\sim 10 \unit{\Myr}$. The overall evolution looks similar to that of the FID model, but it occurs more swiftly, leading to a somewhat shorter lifetime.

\begin{figure}
    \centering
    \includegraphics[width=8cm]{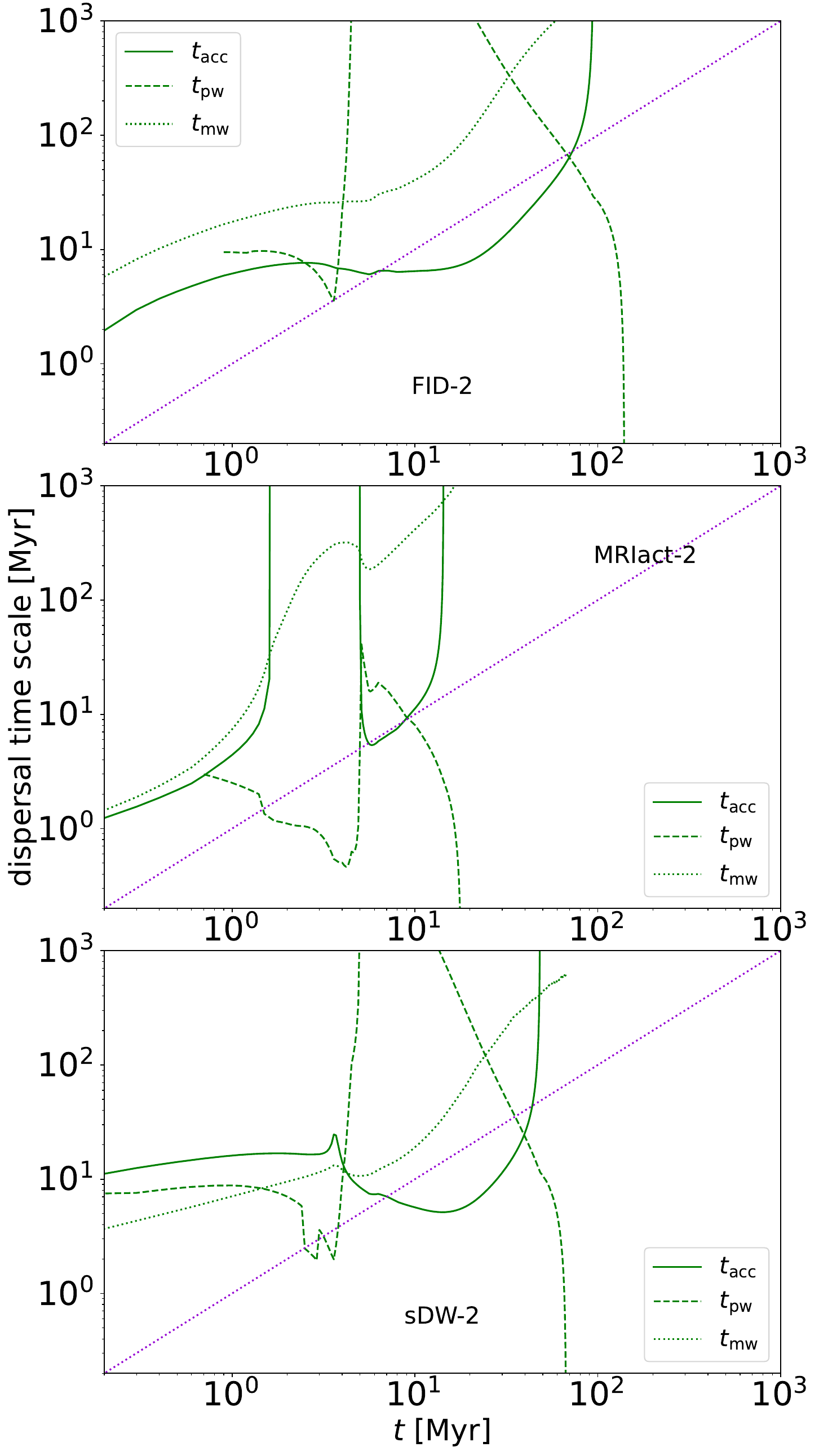}
    \caption{Time evolution of dispersal timescales in models FID-2 (top), MRIact-2 (middle), and STDW-2 (bottom panel). The evolution of $t_{\mathrm{acc}}, t_{\mathrm{pw}}$, and $t_{\mathrm{mw}}$ are presented in the same style as in Figure~\ref{fig:fiducial_t_dispersal}.    
    }
    \label{fig:t_dispersal_AM}
\end{figure}

Figure~\ref{fig:t_dispersal_AM} shows the evolution of dispersal timescales for models FID-2, MRIact-2, and sDW-2. 
The middle panel helps us to interpret the complex evolution of $\Sigma$ in the MIRact-2 model described above. We see that the efficient MRI viscosity causes rapid mass accretion in an early stage of $t \lesssim 1 \Myr$, which is followed by effective photoevaporation until the epoch of $\simeq 4 \Myr$.
The accretion timescale $t_\mathrm{acc}$ diverges at $t \simeq 2 \Myr$ when accretion onto the central star halts due to the clearing of the inner disk. 
This occurs because the inner disk accretes due to viscous angular momentum transport, while mass supply from the outer region 
disperses due to photoevaporation around $10 \unit{au}$, impeding further mass inflow. 
The photoevaporation timescale $t_{\rm pw}$ increases significantly at $\simeq 4 \Myr$, when the host star loses the surface convective zone.
However, since the disk mass is much lower than in FID-2 and the shielding effect of EUV radiation is ineffective, photoevaporation is weakened but $t_{\rm pw}$ does not greatly exceed a few $\times 10 \unit{\Myr}$. 
With the suppression of the photoevaporation that blocks the mass inflow, the inner part of $r \lesssim 100\unit{au}$ is refilled by the mass supply from the outer part. As the disk mass further decreases by mass accretion onto the central star, the photoevaporation returns to be effective, ultimately leading to the quick dispersal of the entire disk.

The MRIact-2 model exhibits the same evolutionary behavior as the ``revenant disks'' described in \cite{2024_Ronco} (compare their Fig.~5 and our Figure~\ref{fig:Sigma_AM}). In these disks, a cavity initially forms but is later refilled after X-ray photoevaporation weakens following the disappearance of the convective zone.

The bottom panel of Figure~\ref{fig:t_dispersal_AM} shows that in the sDW-2 model the MHD disk wind timescale $t_\mathrm{mw}$ is shorter than that in the FID-2 model throughout the evolution.
Therefore, strong mass loss by disk wind makes disk mass 
disperse faster and the photoevaporation timescale $t_\mathrm{pw}$ decreases faster. Therefore, the lifetime of the disk is shorter than in the fiducial model.

\subsection{Modifying Mass Distribution of Initial Disk}
\label{initial_diskmass}

In this section, we examine the effects of modifying the initial disk mass distribution. We vary two parameters: the initial disk mass and the size characterized by the cut-off radius $r_{\mathrm{cut}}$ (see Equation \ref{eq:sgm_ini}).

\begin{figure}
    \centering
    \includegraphics[width=8cm]{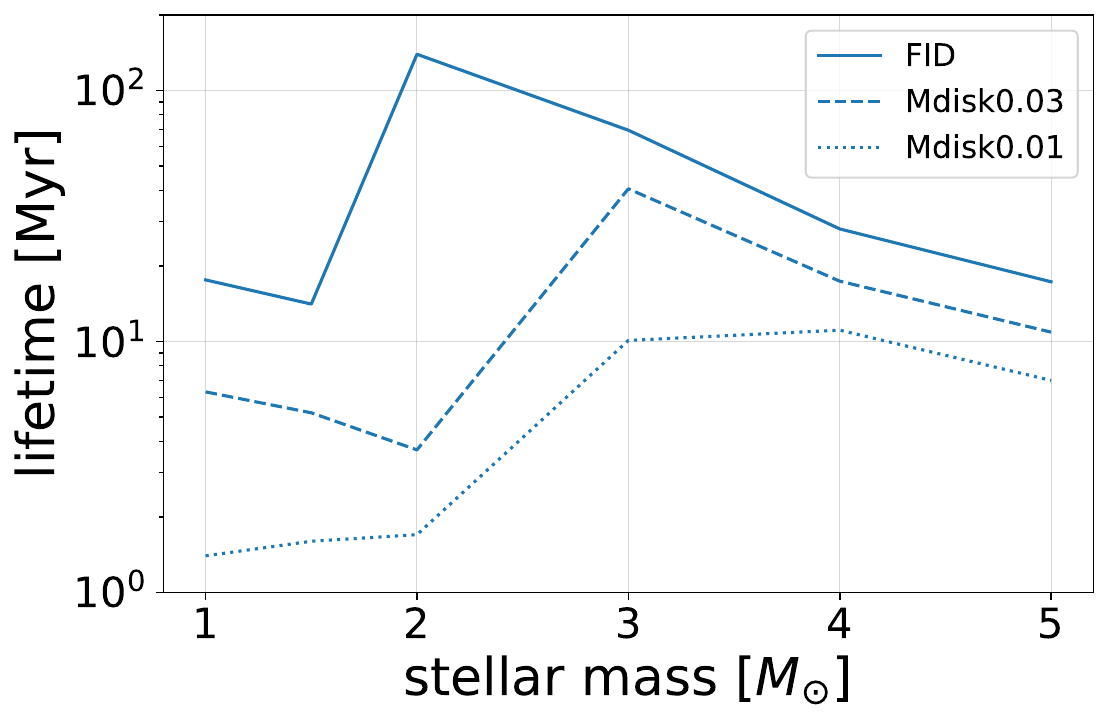}
    \caption{
Effect of decreasing initial disk mass on disk lifetime. The solid, dashed, and dotted lines represent models FID, Mdisk0.03, and Mdisk0.01. 
    }
    \label{fig:lifetime_Mdisk001}
\end{figure}

Figure~\ref{fig:lifetime_Mdisk001} shows how the disk lifetime depends on the host star mass in FID, Mdisk0.01, and Mdisk0.03 models.
FID models assume that the disk starts with a mass equal to 10\% of the mass of the central star, whereas Mdisk0.01 and Mdisk0.03 models employ lower mass ratios of 3\% and 1\%, respectively (refer to Table~\ref{tab:model} for specific model details).
The figure indicates that at a given stellar mass the lifetime is systematically shorter with the lower initial disk mass. In particular, the decrease is prominent for the mass range of $\simeq 1-3 \Msun$. The longest lifetime appears at the higher stellar mass with the lower initial disk mass. For example, the lifetime is the longest at $M_* \simeq 4\Msun$ for model Mdisk0.03, while it is at $\simeq 2\Msun$ for model FID.

\begin{figure*}
    \centering
    \includegraphics[width=\linewidth]{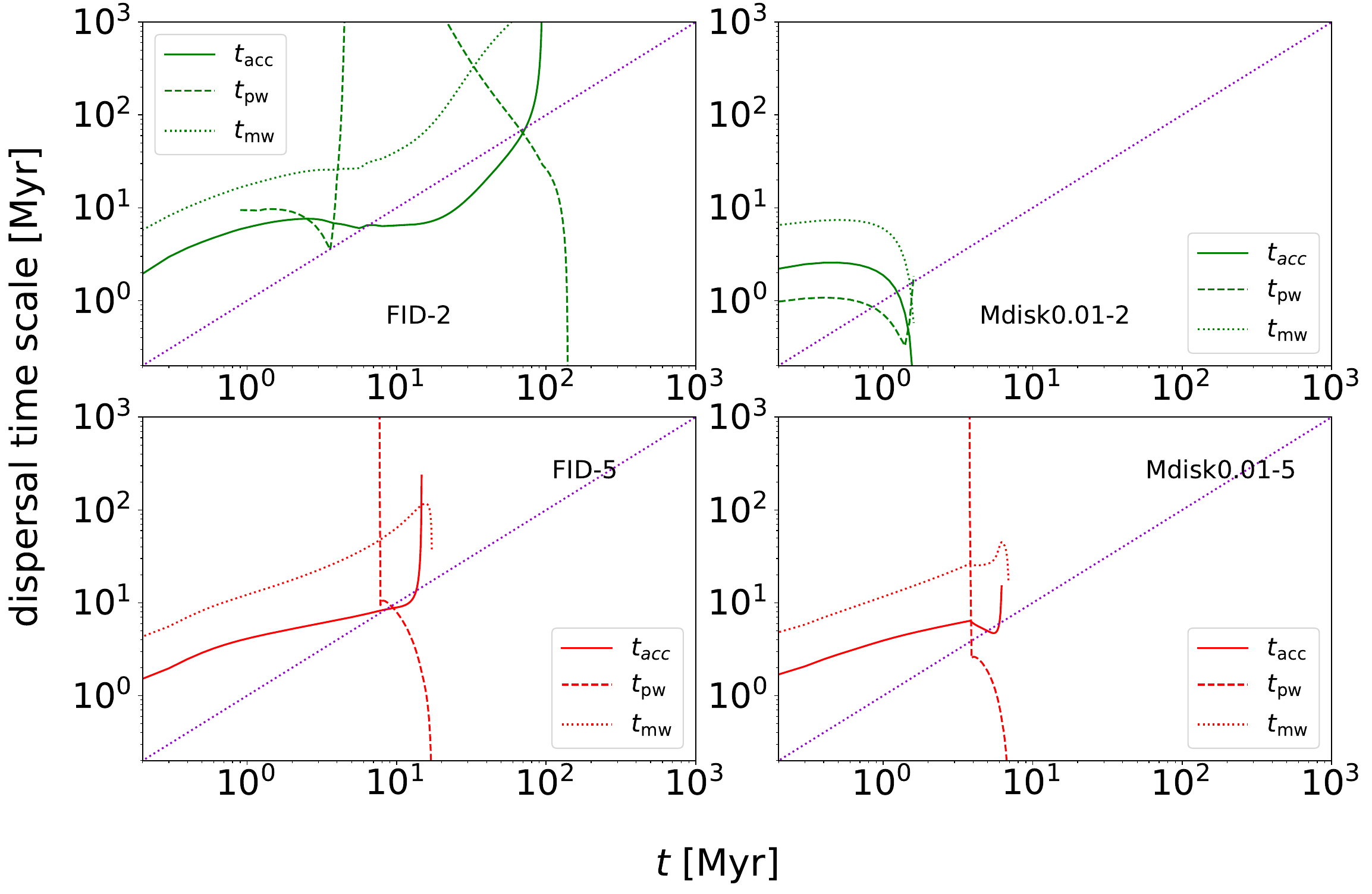}
    \caption{Time evolution of the dispersal timescales for models FID-2 (top-left), Mdisk0.01-2 (top-right), FID-5 (bottom-left), and Mdisk0.01-5 (bottom-right panel). The evolution of $t_{\mathrm{acc}}, t_{\mathrm{pw}}$, and $t_{\mathrm{mw}}$ are presented in the same style as in Figure~\ref{fig:fiducial_t_dispersal}. 
    }
    \label{fig:t_dispersal_Diskmass001}
\end{figure*}

Figure~\ref{fig:t_dispersal_Diskmass001} shows the evolution of the dispersal timescales in FID and Mdisk0.01 models for the specific cases where the host stellar mass is $2$ and $5 \Msun$ (i.e., FID-2,5, and Mdisk0.01-2,5). 
The top right panel shows that in model Mdisk0.01-2 the disk rapidly  
disperses at $\simeq 2 \unit{\Myr}$ by photoevaporation, which is before the host star loses the surface convective layer at $\simeq 4 \unit{\Myr}$ (c.f. top left panel for FID-2 model). The rapid disk dispersal is feasible because $t_\mathrm{pw}$ is significantly less in Mdisk0.01-2 compared to FID-2, attributed to the smaller $M_\mathrm{disk}$. Furthermore, the EUV shielding effect is absent from the beginning in the Mdisk0.01-2 model. 
In contrast, FID-5 and Mdisk0.01-5 models show a similar overall evolution, except that it is more rapid in Mdisk0.01-5 model (lower panels). The EUV shielding effect is available from an early stage up to several Myr. 
The lifetime is mainly determined by the epoch in which the photoevaporation begins to operate as the EUV shielding effect ceases, which is marked by the sudden drop of $t_\mathrm{pw}$.

For the models described above, we have fixed the initial disk size, setting $r_{\mathrm{cut}}=30 \unit{au}$ following \cite{Suzuki+16}.
However, observations suggest positive correlations between the host star mass and the initial disk size. For example, \cite{Weder+2023} used an initial disk model where $r_{\mathrm{cut}} \propto M_*^{0.25}$, while \cite{Komaki+2023} used a model where $r_{\mathrm{cut}} \propto M_*^{1}$. We also study the effect of changing the initial disk size with model Rcut1, for which we assume $r_\mathrm{cut} = 30 \unit{au}~(M_*/\Msun)^{1}$.

\begin{figure}
    \centering
    \includegraphics[width=8cm]{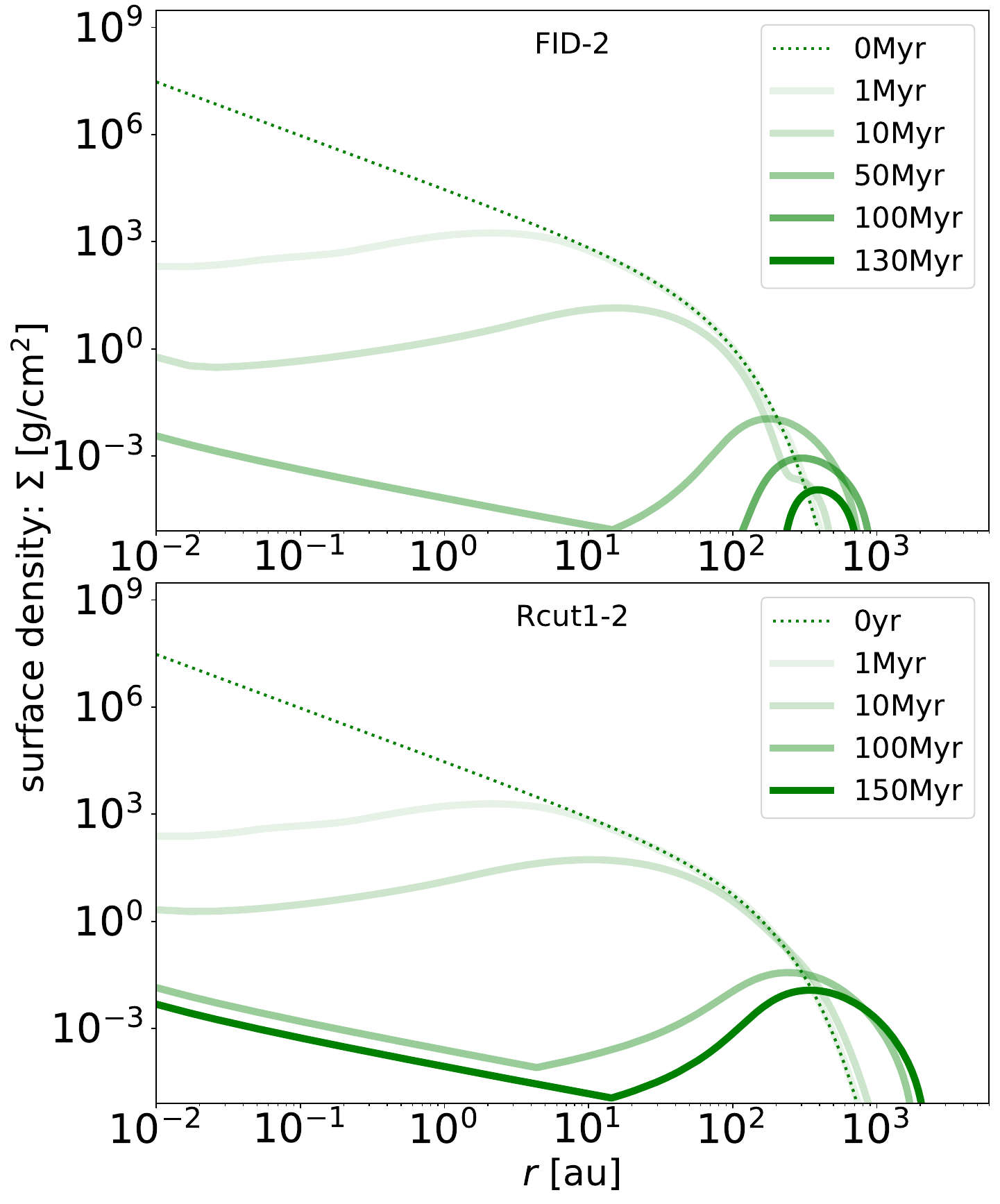}
    \caption{Evolution of the gas surface density for models FID-2 (top) and Rcut1-2 (bottom), where the initial mass distribution cut-off radii are $r_{\mathrm{cut}}=30\unit{au}$ and $60\unit{au}$, respectively (see also Table~\ref{tab:model}). Each panel displays various lines representing snapshots taken at different epochs, as shown in the legend.
    }
    \label{fig:Sigma_rcut}
\end{figure}

Figure~\ref{fig:Sigma_rcut} shows the evolution of the surface density in models FID-2 and Rcut1-2, where $r_\mathrm{cut} = 30 \unit{au}$ and $60\unit{au}$, respectively. 
In Rcut1-2, more disk mass extends to the outer region compared to FID-2. Since the photoevaporation effect is weaker in the outer region, the disk lifetime in Rcut1-2 is extended to more than $150 \unit{\Myr}$ due to the additional mass in the outer disk.

Although we focus here only on the evolution for Rcut1-2, Rcut1 models have a maximum lifetime around $M_* = 2 \Msun$. Therefore, we can explain gas-rich debris disks by the primordial-origin scenario, even if the initial disk size is positively correlated with the host star mass as $r_{\mathrm{cut}} \propto M_*^{1}$.

\section{Discussions}
\label{Discussion}

Our small-grain-depleted disk model predicts generally long lifetimes, often nearing or exceeding $\approx 10 \unit{\Myr}$. 
The lifetimes peak for stars with masses of $M_* = 2$--$3 \Msun$, consistently exceeding $10 \unit{\Myr}$. 
The model also suggests that, essentially, accretion at a detectable level continues as long as the gas disk is present. 
In this section, we analyze our results in more detail to evaluate the validity of our model and assess the plausibility of the primordial-origin scenarios.
In Section~\ref{Plausibility}, we discuss the plausibility of primordial-origin scenario by comparing our results with observation results. In Section~\ref{sec:long-accreting_disks}, we discuss the evolution of accretion rate to understand real disk evolution. In Section~\ref{model_limit}, we discuss our model limitations.

\subsection{On the Plausibility of Primordial-Origin Scenario}
\label{Plausibility}
The most critical necessary condition for gas-rich debris disks to be primordial origin is the existence of a plausible parameter space where protoplanetary disks can survive beyond $10\Myr$.  
The one-zone model of \citet{Nakatani+23} demonstrated that this condition is met if protoplanetary disks are sufficiently depleted of small grains, rendering FUV photoelectric heating ineffective at driving vigorous photoevaporation. 
Their model predicts the longest lifetimes at $M_* = 2$--$3 \Msun$, consistent with the relatively high incidence of gas-rich debris disks around early A~stars (within current detection limits).

\cite{2024_Ronco} also show some of their models incorporating FUV-induced photoevaporation have long lifetimes exceeding 10 Myr. These models confine the disk mass to the area within 10 au. While \cite{2024_Ronco} excluded MHD disk winds, they found that the transfer of angular momentum in their low-viscosity models was less significant than in our models, resulting in a prolonged lifetime for the inner disk. Conversely, the impact of FUV-driven photoevaporation was more pronounced in their models compared to ours, leading to a shorter lifetime for the outer disk than in our models.
The present study has extended the one-zone (0D) model of \citet{Nakatani+23} into a 1D secular evolution model and confirmed the findings of the previous work. 
A new feature introduced by our model for the evolution of small-grain-depleted disks is that the accretion rate remains ongoing as long as the disk persists (Figure~\ref{fig:fiducial_accrate}). 
This implies that if gas-rich debris disks are of primordial origin, ongoing accretion may still be occurring. Therefore, searching for the accretion signatures in debris disks could serve as a potential diagnostic for determining the origin of the gas. 

Currently, ongoing accretion has been reported in only a handful of known gas-rich debris disks, namely HD~141569 \citep{2006_Garcia-Lopez, 2013_Salyk} and 49~Ceti/HD~9672 \citep{2020_Wichittanakom, 2023_Grant}. 
However, given the young age of HD~141569 \citep[$\sim 5\Myr$][]{} and the fact that it is considered to be in an intermediate stage between protoplanetary and debris disk phases (often referred to as a hybrid disk), there is no surprise it retains a detectable, high accretion rate. 

For 49~Ceti, the identification of accretion signature originally comes from the H$\alpha$ emission data of \cite{2018_Vioque}. 
However, other authors have reported that the optical spectroscopy of 49~Ceti exhibits the spectrum of an early A~star without H$\alpha$ emission or UV excess \citep{2007_Carmona}, leaving some uncertainty about the presence of accretion in this system. 

To the best of our knowledge, the accretion status of the other gas-rich debris disks remains unclear based on our review of the literature. While these disks are normally considered non-accreting, there is evidence of possible ongoing accretion at least for debris disks hosted by low-mass stars ($<20\Myr$), as indicated by the detection of \ion{H}{1} lines in Spitzer spectra \citep{2015_Rigliaco}. 
Furthermore, the current sensitivity limit for detecting accretion in Herbig stars is $\sim 10^{-8}\Msun\yr^{-1}$ \citep{Brittain+2023}, which is much higher than the accretion rates predicted by our model. 
This highlights the value of conducting a systematic survey for accretion signatures in gaseous debris disks using tracers with higher sensitivities unless such a survey has not already been undertaken.

We however note that the mere nondetection of accretion signatures in gas-rich debris disks should not be taken as conclusive evidence against primordial-origin scenarios. 
There are potential accretion-inhibiting processes not considered in the present study, such as the presence of inner massive planets \citep{Manara+2019,Manara+23} and disk magnetization \citep{2013_Armitage, 2016_Bai}. 
Their influences should be examined from both theoretical and observational (statistical) perspectives. 

Accretion can generally generate UV and X-ray radiation, which affect disk lifetimes through photoevaporation. In Appendix~\ref{sec:accretion-generated_EUV}, we explore this impact using a simplified model and find that accretion-generated EUV can significantly shorten gas lifetimes for $M_* \approx 2$--$3 \Msun$, although they remain $> 10\Myr$ with our fiducial parameters. While this result is preliminary due to substantial uncertainties in the magnitudes of accretion-generated UV and X-ray radiation, primordial-origin scenarios may overall favor the existence of some accretion-inhibiting processes. 

Another criterion for assessing the plausibility of a primordial-origin scenario could be whether models align with relatively short inner disk lifetimes suggested by H$\alpha$ and infrared disk fractions \citep[a few million years; e.g.,][]{2014_Yasui, Ribas2015}, while simultaneously retaining the outer disk gas traced by CO, \ion{C}{1}, and \ion{O}{1} for $\gtrsim 10\Myr$. 

Our model predicts a significant gas depletion in the inner radii on timescales of a few million years (Figures~\ref{fig:fiducial_Sigma} and \ref{fig:Sigma_AM}) but also suggests a high accretion rate for a longer period, which appears inconsistent with the rapid decline in the H$\alpha$ disk fraction for Herbig stars \citep{2014_Yasui}.

The potential inconsistency between this statistical trend and our prediction could be interpreted as (1) again, there are additional processes that inhibit accretion in gas-rich debris disks, (2) gas-rich debris disks arise from an outlier population of Herbig disks characterized by exceptionally prolonged accretion (see Section~\ref{sec:long-accreting_disks} for further discussion), or (3) the primordial-origin scenario is unlikely. 
Thus, exploring the possibilities (1) and (2) in future studies is crucial to conclusively determining whether gaseous debris disks are indeed secondary in origin, as in the current standard interpretation.

It is important to note that age estimation and sample selection techniques for clusters significantly impact inner disk lifetime estimates based on disk fractions. 
\citet{2018_Richert} demonstrated that using the magnetic pre-main-sequence star models of \citet{2016_Feiden} results in longer estimated lifetimes, by about a factor of a few, compared to the canonical estimates (a few million years). 
Similarly, \citet{Pfalzner_2022} found that focusing solely on nearby clusters leads to longer lifetime estimates (5--10$\Myr$) by avoiding the over-representation of higher-mass stars in distant clusters due to limiting magnitudes \citep[See also][for longer lifetime estimates]{2024_Pfalzner}.

These systematic uncertainties should be considered when assessing primordial-origin models by comparison with inner disk lifetimes.

\subsection{Long Accretors beyond 10 Myr}    
\label{sec:long-accreting_disks}
From a broader perspective, beyond just gaseous debris disks, understanding the origin of long-lasting accretors ($\geq 10 \Myr$-old) is key to comprehending disk evolution. 
Long accretors are observed across a wide range of host stellar masses. 
High accretion rates have been found for Herbig stars even at estimated ages of around $10\Myr$ and beyond \citep{2019_Arun}. 
As mentioned in the previous section, a signature of ongoing accretion has been detected for debris disks ($<20\Myr$) hosted by low-mass stars through \ion{H}{1} lines \citep{2015_Rigliaco}.
Similarly, \citet{2020_Manara} identified high accretion rates for low-mass stars in the Upper Sco star-forming region ($5$--$10\Myr$), which are comparable to those seen in younger star-forming regions. 
Notably, accreting M~dwarfs with ages of $> 10\Myr$ are referred to as Peter Pan disks \citep[e.g.,][]{2016_Silverberg, 2020_Silverberg}. 

\begin{figure*}[htbp]
    \centering
    \includegraphics[clip, width = 0.49\linewidth]{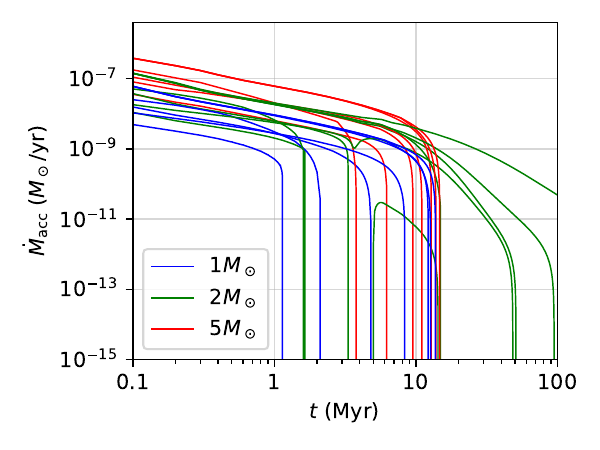}
    \includegraphics[clip, width = 0.49\linewidth]{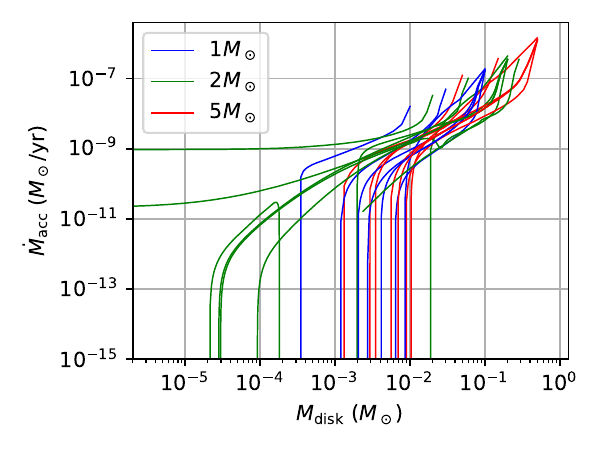}
    \caption{
    (left): Time evolution of accretion rates for various models (Table~\ref{tab:model}) with stellar masses of $1 \Msun$ (blue), $2 \Msun$ (green), and $5 \Msun$ (red). 
    Accretion rates for the other stellar masses are omitted for clarity, but they show intermediate evolutionary trends between those depicted here. (right) Corresponding evolutionary paths on the $\Mdisk$--$\Macc$ map.
    }
    \label{fig:fiducial_accrate}
\end{figure*}
The origin of long accretors remains unclear, but our small-grain-depleted disk scenario offers a possible explanation for their longevity.  
The left panel of Figure~\ref{fig:fiducial_accrate} shows the time evolution of accretion rates for various models explored in the present study (Table~\ref{tab:model}).
Accretion consistently lasts beyond $10\Myr$ regardless of the stellar mass, except in models with lower initial disk masses (Mdisk0.01 and Mdisk0.03) or high turbulent stress (MRIact). 
This suggests the trend that relatively massive disks ($\approx 0.1 M_*$) naturally evolve into long accretors, unless turbulent stress is particularly strong ($\alpha \sim 10^{-2}$) throughout the disk.
Recent ALMA observations suggest much lower $\alpha$ values \citep[$10^{-4}$--$10^{-3}$; ][for recent review]{Manara+23}, meaning that disks following the evolutionary path of our MRIact models are likely rare. 
This implies that long accretors originate from a population of disks with large initial masses. 

The longest accretion is found for $M_* = 2\Msun$ stars (up to $\approx30$--$50\Myr$), with accretion rates of $\Macc \sim 10^{-11}$--$10^{-10}\Msun\yr^{-1}$. For $M_* = 1$ and $5\Msun$ stars, it lasts up to approximately $15\Myr$.
The relatively slow decay in $\Macc$ for $M_* = 2\Msun$ stars implies that their disks retain low disk masses for a relatively long time.
This is indicated by the right panel of Figure~\ref{fig:fiducial_accrate}, where $\Mdisk$ is $\sim 10^{-4}$--$ 10^{-3}\Msun$ for $M_* = 2\Msun$ when the accretion rates drop to $\Macc \sim 10^{-11}$--$10^{-10}\Msun\yr^{-1}$ at $t > 10\Myr$. 
Assuming the interstellar elemental abundance of carbon ($\sim 10^{-4}$), the corresponding CO mass during this phase is simply estimated to be $M_\mathrm{CO} \sim 0.06$--$0.6 \,M_\oplus$. 
This is comparable to or higher than the observed mass of the most massive gas-rich debris disks \citep[e.g., HD~21997; $M_\mathrm{CO}\approx 0.06\,M_\oplus$][]{Kospal2013}.
However, note that our estimated CO mass represents the upper limit, as photodissociation would likely reduce the actual $M_\mathrm{CO}$ to values closer to those observations.

Our models with $M_* = 1\Msun$ show accretion rates consistent with those obtained for the Upper Sco disks by \citet{2020_Manara}. 
However, our models predict significantly higher disk masses compared to observational values, which are estimated by $100\times$ the dust disk masses. 
This discrepancy could imply that additional dispersal processes not included in this study, like external photoevaporation, play a significant role, or that the actual dust-to-gas mass ratio is much lower than the standard assumption \citep{2020_Sellek}, as pointed out by the original paper, \citet{2020_Manara}. 

To summarize, we propose that long accretors originate from initially massive disks ($\sim 0.1 M_*$), where PAHs and very small grains become sufficiently depleted during the disk evolution, rendering internal and external FUV-driven photoevaporation ineffective; otherwise, disks are unlikely to survive for $\gtrsim10\Myr$ based on the results of previous studies \citep{Komaki+2023, 2024_Ronco}.
To further investigate the origin of long accretors, incorporating grain growth in the 1D model is essential for accurately determining the dust-to-gas mass ratio and the degree and likelihood of small-grain depletion.

\subsection{Model Limitations}
\label{model_limit}
\paragraph{Longest-living population}
As noted in Section~\ref{method}, this study has focused on the evolution of disks where small grains are depleted, and which have relatively high initial masses, as these disks are presumed to be the longest-lived and therefore optimal for examining the plausibility of primordial-origin scenarios. It is important to emphasize that the results presented here may not be either typical or representative for protoplanetary disk's evolution. Our model merely demonstrates how protoplanetary disks could achieve long lifetimes. The likelihood of a protoplanetary disk following this evolutionary pathway heavily depends on the uncertain probability of small-grain depletion and needs to be assessed through population synthesis, taking into account the effects of grain growth. 

\paragraph{Depletion of Very Small Grains and PAHs}
We have neglected FUV photoevaporation in our model based on the assumption of very-small-grain depletion. However, \cite{2024_Ronco} and \cite{Kunitomo+21} show that FUV photoevaporation becomes significant at approximately 1--3~$\Myr$ for stars with masses of 2--3$M_{\odot}$, when their FUV luminosities increase due to stellar evolution. Even in such cases, FUV photoevaporation can remain negligible if very-small-grain depletion occurs sufficiently early, before the luminosity increase. The threshold abundance of PAHs and very small grains below which FUV photoevaporation becomes ineffective remains uncertain, but we estimate it to be approximately 0.1–1\% of the ISM level \citep{Nakatani_2018a,Nakatani+18}.

There are observational evidence suggesting depletion of PAHs and small grains. PAHs have been detected only $\sim$10 \% of T Tauri stars \citep{Geers_2006,Geers_2007,Oliveira_2010,Vicente_2013} and 70\% of Herbig stars \citep{Acke_2010}. 
Additionally, mid-IR spectra and IR scattered light imaging of PPDs also imply the depletion of small grains (approximately $\lesssim 2\unit{\mu m}$) in the disk atmosphere \citep[e.g.][]{Furlan_2006,Furlan_2011,Mulders_2013}.

Several processes can reduce the abundance of very small grains. Grain growth combined with radial drift is known to remove dust from the disk \citep{Sellek_2020}, and \cite{Pinilla_2022} show that the radial drift is notably fast around intermediate-mass stars. In addition, radiative forces exerted by intermediate-mass stars preferentially deplete small dust grains \citep{Owen_2019,Nakatani2021}.

Even considering these effects, some studies suggest that small grains can persist for several million years \citep{Birnstiel_2011,Drazkowska_2016,Drazkowska_2017,Drazkowska_2021,Stammler_2023,Guilera_2020}. However, in such cases, small grains survive only in specific regions of the disk.
For example, they may be concentrated in the inner disk, leading to depletion in the outer regions \citep{Drazkowska_2016,Drazkowska_2017}, or accumulate locally in pressure bumps \citep{Guilera_2020}. 
While FUV photoevaporation can contribute to local gas dispersal, it is unlikely to disperse the entire disk. In addition, these focus on the micron-size grains made of the silicate- and ice-rich materials \citep{Furlan_2006}. The photoelectric effect is mainly contributed by the nm-sized grains made of carbon-rich materials \citep{Bouteraon_2019,Preibisch_1993} which likely have sticking and fragmentation properties differing from micron-size grains. Indeed, models have been proposed that demonstrates the rapid dissipation of carbon grains \citep{Okamoto_2024}. Therefore, our model does not contradict these studies on dust evolution. A more detailed quantitative analysis incorporating grain growth and FUV photoevaporation is necessary, which we leave for future studies.

\paragraph{Planetesimal/planet formation}
In addition to grain growth, the formation of planetesimals and planets could significantly affect disk evolution, especially as the dust surface density decreases \citep{Drazkowska_2023}. These bodies may dynamically alter the gas disk evolution, beyond the predictions of this study, by creating gaps/cavities, forming nonaxisymmetric structures, inhibiting gas accretion towards the star, and releasing gases through collisions. Incorporating these effects is essential for understanding the origins of gaseous debris disks and enabling more detailed comparisons with observations.

\paragraph{Environmental effects} 
We have not included external photoevaporation in our model, as our focus is on gaseous debris disks that are detected nearby environments without strong external radiation. Additionally, external FUV photoevaporation is likely ineffective for small-grain-depleted disks. However, it is important to note that our model does not account for external photoevaporation when applying our results to general systems. Within the \ion{H}{2} regions near massive stars, external {\it EUV} photoevaporation could significantly shorten disk lifetimes, even in small-grain-depleted disks.

More broadly, late-stage infall from the interstellar medium \citep[e.g.,][]{Dullemond+19} is also an important process. It has the potential to extend the gas disk lifetime by replenishing its material \citep{Winter+24}. Considering this effect would also be crucial for exploring the origins of long-lived gas disks in general. 

\paragraph{Chemistry}
In primordial-origin scenarios, the gas disk is assumed to be hydrogen-dominated, similar to protoplanetary disks. In contrast, gaseous debris disks are typically detected through emission lines of metal species, namely CO, \ion{C}{1}, \ion{C}{2}, and \ion{O}{1}. Consequently, thermochemistry modeling is necessary to directly compare our results with observations. This modeling can help determine whether these tracers can remain detectable above observational limits and assess their abundance relative to hydrogen. 

Such investigations are also important for evaluating whether our model can account for the findings from recent deep ALMA observations \citep{2022_Smirnov-Pinchukov}. While disks around relatively old ($8$--$10\Myr$) Herbig~Ae stars show molecular emissions of C-bearing species like \ce{HCO+} and \ce{C2H}, these emissions are not detected in gaseous debris disks. Interestingly, both groups exhibit similar amounts of CO. Combining thermochemistry modeling with the time-evolving picture from our model would offer valuable insights into understanding these characteristics.

\section{Conclusions}
\label{conclusion}
We have performed 1D secular disk evolution simulations for small-grain-depleted disks, incorporating self-consistent descriptions of dispersal processes. These simulations aimed to assess the plausibility of the primordial-origin scenario for gas-rich debris disks. Our key findings are summarized as follows:

\paragraph{1.} Gas disk lifetimes exceed $10 \Myr$, peaking at $M_* = 2 \Msun$ (Figure~\ref{fig:lifetime_fiducial}), which aligns with the previous one-zone model by \citet{Nakatani+23}. This result is consistent across a wide variety of disk evolution models (Table~\ref{tab:model}), including those where the shielding effect of inner disk winds is ignored, or where strong disk winds are incorporated (Figures~\ref{fig:lifetime_Noshield} and \ref{fig:lifetime_all_AM}). 
The exception occurs when the initial disk mass is relatively small ($\lesssim 0.01 M_*$; Figure~\ref{fig:lifetime_Mdisk001}), or when the turbulent stress within the disk is relatively strong ($\overline{\alpha_{r\phi}} > 10^{-2}$; green line of Figure~\ref{fig:lifetime_all_AM}). In these cases, protoplanetary disks are less likely to survive beyond $10\Myr$. 

To summarize, these outcomes demonstrate the plausibility of the primordial-origin scenario, consistent with the relatively high incidence of gas-rich debris disks around early A-type stars. If these gas-rich debris disks are remnants of PPDs, they likely originate from initially massive ($\sim 0.1M_*$) disks with minimal turbulent viscosity ($\overline{\alpha_{r\phi}} \lesssim 10^{-3}$).

\paragraph{2.} In disks around $2 \Msun$ stars, gas remains at $\sim 10 - 1000\unit{au}$ (Figures~\ref{fig:fiducial_Sigma} and \ref{fig:Sigma_AM}). These ring-like structures are formed by inner disk dispersal due to accretion and the prolonged photoevaporation dispersal timescale in the outer disk, caused by the disappearance of the stellar surface convective layer. In particular, in MRI inactive models, the disk mass remains around $10 - 100\unit{au}$, and these characteristics are consistent with observations. 

\paragraph{3.} While our models predict a rapid surface density decline at inner radii within several million years, potentially consistent with the infrared lifetimes of PPDs (Appendix~\ref{sec:IR_lifetime}), they also suggest continued accretion beyond $10\Myr$ (Sections~\ref{Plausibility} and \ref{sec:long-accreting_disks}). Although debris disks are generally believed non-accreting, there are evidences suggesting that accretion may be occuring in several debris disks (Section \ref{sec:long-accreting_disks}). Given that the accretion status of the known gas-rich debris disks remains unclear, searching for accretion signatures could provide valuable insights into their origins. However, this study has not accounted for effects that could inhibit accretion, such as the presence of a substellar companions, so further exploration is necessary for more meaningful quantitative comparisons. 

Additionally, we stress that beyond gaseous debris disks, our model provides a possible explanation for the presence of long-living accreting disks detected across a wide range of stellar masses, including Peter Pan disks. Previous studies suggest that the origin of Peter Pan disks lies in low-viscosity PPDs \citep[e.g.][]{Coleman+2020}.

\vspace{0.2cm}
In this study, we have focused on protoplanetary disks with small grain depletion and relatively high initial masses, as these conditions are presumed to extend the disks' lifetimes, making them prime candidates for evaluating primordial-origin scenarios. It is crucial to recognize that the outcomes presented here may not necessarily be typical or universally applicable to all protoplanetary disk evolution processes. Our models serve to illustrate possible scenarios in which these disks might achieve prolonged longevity. The likelihood of any given protoplanetary disk following this specific evolutionary path depends significantly on the uncertain factor of small-grain depletion, a phenomenon that must be further explored through population synthesis that also accounts for grain growth.

\renewcommand{\thelinenumber}{}
\begin{acknowledgments}

The authors thank Sebasti\'an Marino, Ilaria Pascucci, Geoffrey Bryden, Munetake Momose, and Taichi Uyama for fruitful discussions on accretion signatures in debris disk systems. We are also grateful to Rolf Kuiper, Takeru Suzuki, and Takahiro Tanaka for their insightful comments on general matters.
This research could never be accomplished without the support by Grants-in-Aid for Scientific Research (TH:19KK0353, 22H00149) from the Japan Society for the Promotion of Science. This work was also financially supported by ISHIZUE 2024 of Kyoto University. 
R.N. and H.M. have been supported by the Japan Society for the Promotion of Science (JSPS), Overseas Research Fellowship.

\end{acknowledgments}

\software{ 
Numpy \citep{numpy}, Matplotlib \citep{matplotlib}, scipy \citep{scipy}
          }

\appendix

\section{Impact of Accretion-Generated EUV Radiation}
\label{sec:accretion-generated_EUV}

\begin{figure}
    \centering
    \includegraphics[width=8cm]{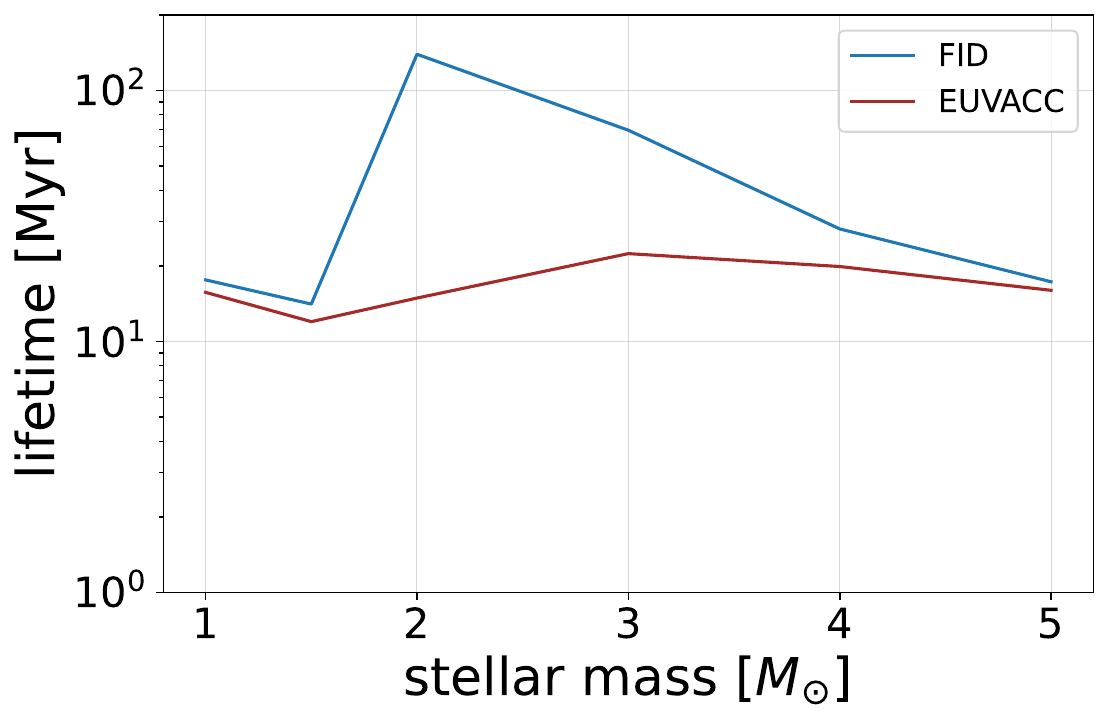}
    \caption{
    Effect of EUV radiation generated by mass accretion $\Phi_{\mathrm{EUV, acc}}$ on the disk lifetime. 
    The green line shows the stellar-mass dependence of the disk lifetime of EUVACC models (see Table~\ref{tab:model}), while the blue line represents our fiducial (FID) models.
    }
    \label{fig:lifetime_YesPhiacc}
\end{figure}

Here we investigate the effect of EUV radiation generated by mass accretion $\Phi_{\mathrm{EUV, acc}}$ in addition to $\Phi_\mathrm{EUV}$ given by Equation \eqref{eq:phiev}. We consider model EUVACC, where 4\% of the gravitational energy is converted to EUV radiation energy as in \cite{Nakatani+23}.
Figure~\ref{fig:lifetime_YesPhiacc} shows the stellar mass dependence of the disk lifetime for EUVACC models. The lifetime is nearly constant at $\simeq 10 \unit{\Myr}$ regardless of the different stellar mass. In particular, the lifetime substantially decreases in the mass range of $2-3 \Msun$ compared to FID models. 

\begin{figure}
    \centering
    \includegraphics[width=8cm]{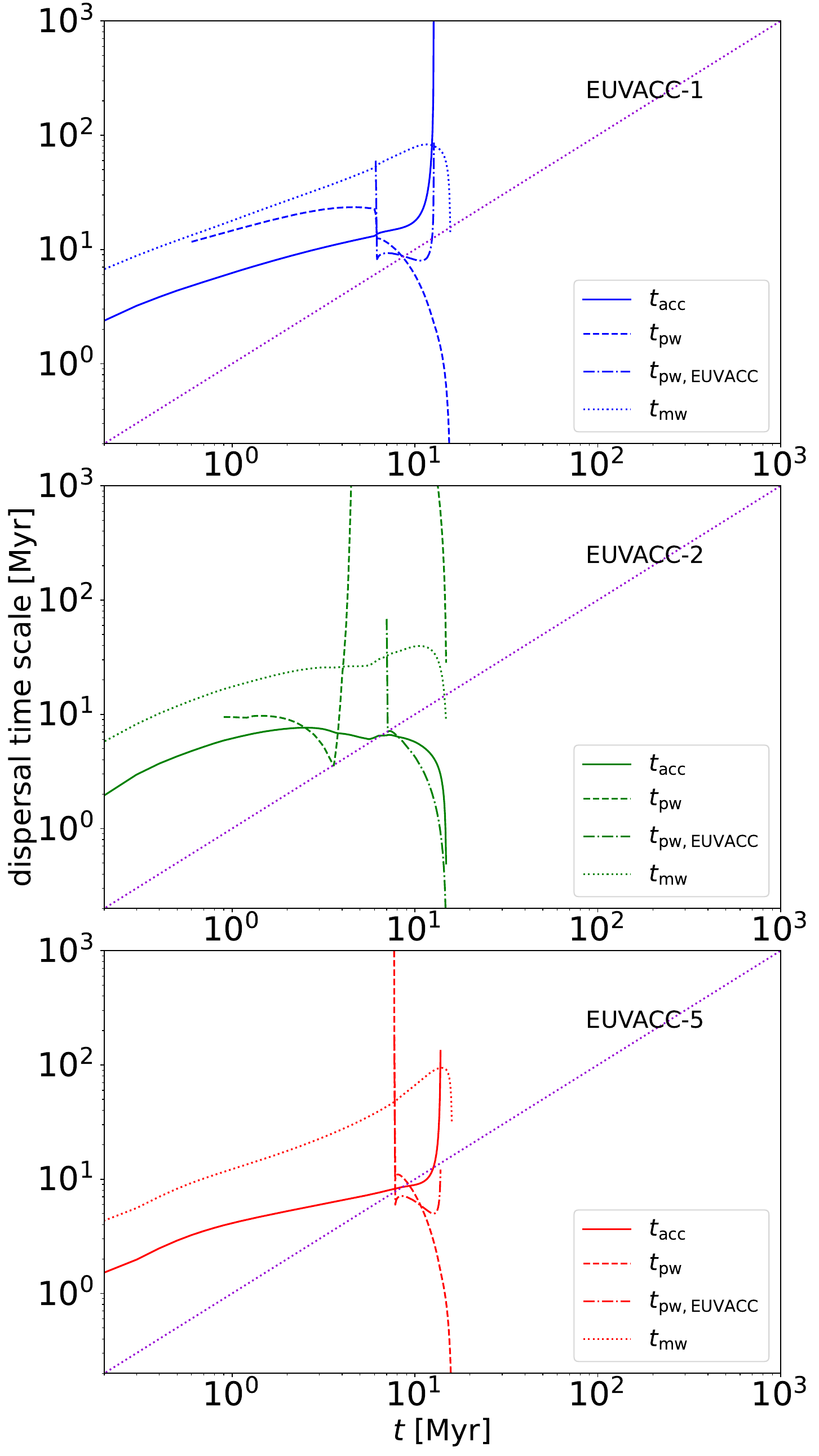}
    \caption{
    Time evolution of the dispersal timescales for models  EUVACC-1 (top), EUVACC-2 (middle), and EUVACC-5 (bottom). The evolution of $t_{\mathrm{acc}}, t_{\mathrm{pw}}$, and $t_{\mathrm{mw}}$ are presented in the same style as in Figure~\ref{fig:fiducial_t_dispersal}. 
    In addition to these, the dispersal timescale representing EUV photoevaporation induced by mass accretion $t_{\mathrm{pw, EUVACC}}$ is represented by the dot-dashed line. 
    }
    \label{fig:t_dispersal_YesPhiacc}
\end{figure}

Figure~\ref{fig:t_dispersal_YesPhiacc} shows the evolution of the dispersal timescales in EUVACC models. The features shown in Figure~\ref{fig:lifetime_YesPhiacc} can be understood by considering balance between the timescales $t_{\mathrm{pw}}$ and $t_{\mathrm{pw, EUVACC}}$, where the latter is defined as 
\begin{equation}
t_{\mathrm{pw, EUVACC}} = \frac{M_\mathrm{disk}(t)}{\dot{M}_\mathrm{pw, EUVACC} (t)} ,
\end{equation}
where $\dot{M}_{\mathrm{pw, EUVACC}}$ is the total mass loss rate integrated over the whole disk caused only by the accretion-generated EUV radiation.
In the middle panel of Figure~\ref{fig:t_dispersal_YesPhiacc}, EUVACC-2 model reveals that the timescale $t_{\mathrm{pw, EUVACC}}$ is significantly shorter than $t_{\mathrm{pw}}$, suggesting that EUV radiation from accretion considerably enhances photoevaporation. Once the surface convective layer vanishes around $t \simeq 4\Myr$, the supplementary term $\Phi_{\mathrm{EUV, acc}}$ becomes the primary driver of photoevaporation, overtaking the other factors in $\Phi_\mathrm{EUV}$. Consequently, the disk lifetime is greatly reduced compared to the FID-2 model. On the other hand, the top and bottom panels of Figure~\ref{fig:t_dispersal_YesPhiacc} show that for the EUVACC-1 and 5 models, $t_{\mathrm{pw, EUVACC}}$ is nearly comparable to $t_{\mathrm{pw}}$. In these models, the absence of a sharp decline in $\Phi_\mathrm{EUV}$ means that $\Phi_{\mathrm{EUV, acc}}$ only marginally boosts photoevaporation, resulting in similar disk lifetimes between FID and EUVACC models for stellar masses of $M_{*}=1$ and $5 \Msun$.

\section{Disk Lifetime inferred by near-infrared emission}
\label{sec:IR_lifetime}

We have estimated the disk lifetime in our 1D model with an epoch when the disk mass becomes nearly zero. In this section, we discuss the lifetime inferred by near-infraraed (NIR) emission, which has often been used in many observations and is sensitive to the presence of an inner part ($\lesssim 10\unit{au}$) of the disk. 

Following \citet{Kimura_2016} and \citet{Kunitomo+21}, we calculate the inner disk lifetimes by analyzing the surface density and temperature distributions. The NIR-emitting region is defined as the region where the temperature exceeds 300 K. Assuming that the disk contains minimal amounts of dust grains, we set the opacity in the NIR to $\kappa_\mathrm{NIR} = 1\unit{cm^{2}/g}$. We define the inner disk lifetime as the epoch when the optical depth in the NIR-emitting region $\kappa_\mathrm{NIR} \Sigma$ falls below unity. Reducing the criterion by an order of magnitude results in approximately a $50\%$ increase in the inner disk lifetime in FID-2. On the other hand, raising the criterion leads to a roughly $30\%$ decrease in the inner disk lifetime in FID-2.

\begin{figure}
    \centering
    \includegraphics[width=8cm]{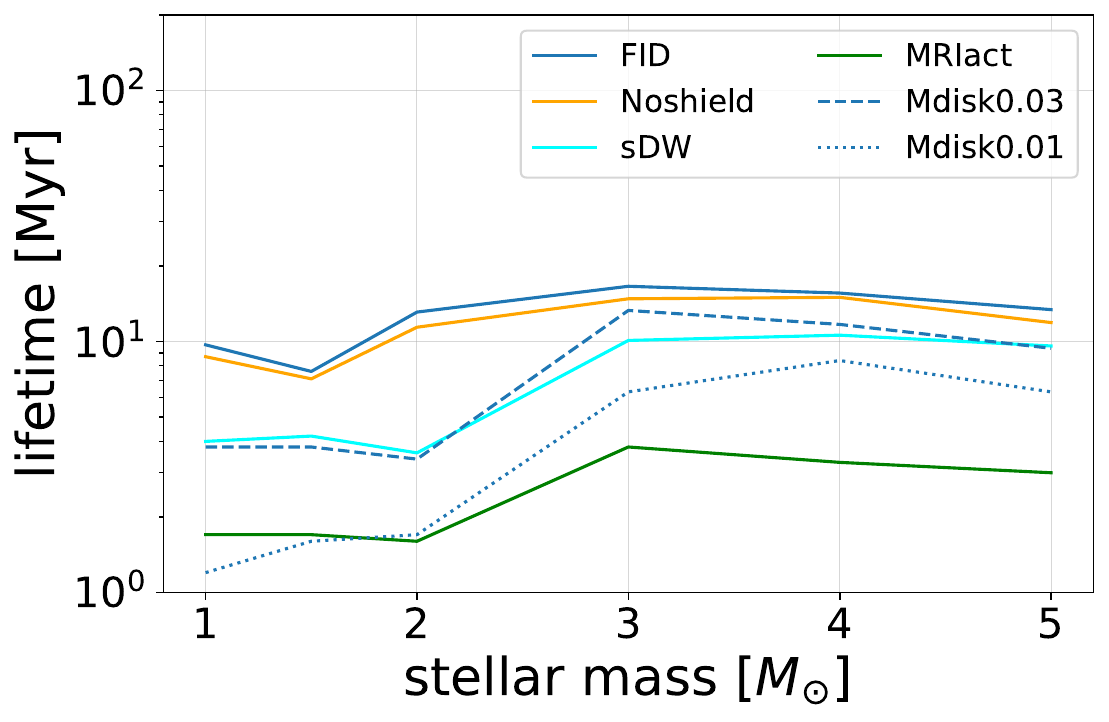}
    \caption{
Dependence of the disk lifetimes on stellar mass when the lifetime is inferred by NIR emission (see text).
    The blue, orange, cyan, and green solid lines represent FID, Noshield, sDW, and MRIact models. The blue dashed and dotted lines represent Mdisk0.03 and Mdisk 0.01 models, respectively.
    }
    \label{fig:lifetime_IR}
\end{figure}

Figure~\ref{fig:lifetime_IR} shows the stellar-mass dependence of the lifetime of the inner disk in different models. The lifetime only weakly depends on the mass of the central star, which differs from our discussion in the main part, where we have defined the lifetime as the epoch when the total disk mass becomes nearly zero. The prominent lifetime peak at about $2 \Msun$ seen in Figure~\ref{fig:lifetime_fiducial} (also noted in Figures~\ref{fig:lifetime_Noshield} and \ref{fig:lifetime_all_AM}) does not appear in Figure~\ref{fig:lifetime_IR}. \citet{Ribas2015} observationally study the stellar-mass dependence of the disk lifetimes inferred by the NIR emission, arguing that the inner disk lifetime with $M_* > 2 \Msun$ is up to two times shorter than that with the less massive stars. Explaining the stellar-mass dependence of the inner disk lifetime as suggested by \citet{Ribas2015} is beyond our scope, since we only consider the extreme cases where the FUV photoevaporation is assumed to be inefficient. More models with $M_* < 1 \Msun$ are also necessary for effective comparisons with the observations. Nevertheless, Figure~\ref{fig:lifetime_IR} may indicate the extent to which currently ignored processes (e.g. FUV photoevaporation) and their stellar mass dependence will be necessary to match observations. As mentioned in the main text, the strength of FUV photoevaporation depends on the number of small grains that survive long in a PPD, which is uncertain. The stellar mass dependence of this effect is also uncertain and should be studied in detail in future work.

\bibliography{reference}

\end{document}